\documentclass[hidelinks,man,floatsintext]{apa7}
\usepackage[american]{babel} 
\usepackage{csquotes}  

\usepackage[style=apa,
backend=biber,
sortcites=true,
sorting=nyt,
maxcitenames=1,
mincitenames=1,
maxbibnames=20]{biblatex}
\DeclareLanguageMapping{american}{american-apa}

\usepackage{times} 

\usepackage[symbol]{footmisc} 

\usepackage[utf8]{inputenc}
\usepackage{microtype}
\usepackage{amsthm}
\usepackage{amsmath}  
\usepackage{amssymb}  
\usepackage{comment}  
\usepackage{lipsum}
\usepackage{lmodern}
\usepackage{iftex}
\usepackage[T1]{fontenc}
\usepackage{textcomp} 
\usepackage{pdfpages}  
\usepackage{bookmark} 

\usepackage{array}
\usepackage{longtable}
\usepackage{lscape}  
\usepackage{rotating}  
\usepackage{multirow}		
\usepackage{tabularx}		
\usepackage{booktabs}  
\usepackage[flushleft]{threeparttable}	
\usepackage{threeparttablex}  

\makeatletter
\newcommand\LastLTentrywidth{1em}
\newlength\longtablewidth
\newcommand{\getlongtablewidth}{\begingroup \ifcsname LT@\roman{LT@tables}\endcsname \global\longtablewidth=0pt \renewcommand{\LT@entry}[2]{\global\advance\longtablewidth by ##2\relax\gdef\LastLTentrywidth{##2}}\@nameuse{LT@\roman{LT@tables}} \fi \endgroup}
\makeatother

\title{Accounting for CAT-Induced Dependency in Differential Item Functioning Detection: A Multilevel Modeling Framework}  
\shorttitle{Computerized Adaptive Testing}

\author{Dandan (Danielle) Chen Kaptur\textsuperscript{1}\footnote[1]{Corresponding author. Email: \href{mailto:danielle.chen@pearson.com}{\nolinkurl{danielle.chen@pearson.com}}.}, Justin Kern\textsuperscript{2}, Chingwei David Shin\textsuperscript{3}, Jinming Zhang\textsuperscript{2}}
\affiliation{
\vspace{0.5cm}
\textsuperscript{1} Pearson\\
\textsuperscript{2} University of Illinois Urbana-Champaign\\
\textsuperscript{3} American Board of Psychiatry and Neurology}

\abstract{Differential item functioning (DIF) detection is an important yet understudied problem in computerized adaptive testing (CAT). In this article, we proposed a two‑level logistic model to improve DIF detection in CAT by explicitly accounting for nuisance effects arising from CAT-induced structural dependency. First, we conceptualized that adaptive item selection induces systematic dependencies among examinees and items through provisional ability estimates, whereas traditional single‑level DIF methods assume independent observations and may yield misleading results in CAT settings. Then, using a numeric example and Monte Carlo simulations, we compared our proposed two-level model with competing single‑level models under various CAT conditions, manipulating test length, exposure control, ability estimator, DIF type, and DIF prevalence. Item‑level Type‑I error and statistical power conditional on joint model convergence were reported for each model. We showed that the proposed two‑level model has improved control of spurious DIF and competitive power relative to single‑level models, particularly with shorter tests and smaller exposure rates. However, we observed that the model convergence varied systematically across simulated conditions, highlighting that inferential accuracy and convergence reliability are intertwined in complex CAT DIF settings. Through this study, we underscored both the promise of multilevel DIF modeling in CAT and the need for future research to jointly evaluate convergence and inferential performance when assessing DIF models.}


\begin{document}

This is a preprint. The Version of Record of this article is published in \textit{Applied Psychological Measurement}, and
is available online at \url{https://doi.org/10.1177/01466216261451502}.

\begin{titlepage}

\textbf{Acknowledgments}\\
    The authors want to express sincere appreciation and gratitude to many people who provided helpful feedback on early drafts of this manuscript, including the journal editor Dr. John R. Donoghue and anonymous reviewers. Also, the authors appreciate the support and opportunities to present preliminary findings at national conferences, from Pearson and University of Illinois Urbana-Champaign.\\

\vspace{10pt}
\textbf{Declaration of conflicting interest}\\
    The authors report there are no competing interests to declare.\\

\vspace{10pt}
\textbf{Funding statement}\\
    This work was supported by the Conference Travel Grant at the College of Education, University of Illinois Urbana-Champaign.\\

\vspace{10pt}
\textbf{Ethical approval and informed consent statements}\\
    Ethical approval and/or written informed consent are not needed for this research because its data are either publicly accessible or simulated via software programs.\\

\vspace{10pt}
\textbf{Data availability statement}\\
    The data that support this research are openly available at \textit{TIMSS 2019 International Database} at https://timss2019.org/international-database/.

\end{titlepage}

\maketitle

\section{Introduction}

Computerized adaptive testing (CAT) has been widely used since the 1980s \parencite{wainer_computerized_2000}. Comparatively less attention has been devoted to the problem of differential item functioning (DIF) in CAT. DIF detection is central to test fairness \parencite{hambleton_fundamentals_1991}, yet most existing DIF methods have been developed for fixed‑form tests that assume all examinees respond to the same items. They include the following: the Mantel-Haenszel procedure \parencite[MH;][]{holland_alternate_1985}, logistic regression procedure \parencite[LR;][]{swaminathan_detecting_1990}, SIBTEST \parencite{shealy_model-based_1993}, and IRT-based likelihood ratio test \parencite[IRT-LR;][]{thissen_beyond_1986}. They have been shown to be ineffective for DIF detection in CAT \parencite[e.g.,][]{nandakumar_catsib_2001,lei_comparing_2006}. 
\\
In CAT, provisional ability estimates influence which items are selected next. We argue that traditional single‑level DIF methods---which assume independent observations---may yield inaccurate results unless adapted for this structure, because the adaptive item‑selection mechanism creates systematic dependencies among items and examinees through provisional ability estimates. It motivates us to explore whether multilevel modeling, known for providing more accurate estimates by accounting for dependencies in hierarchical data \parencite{raudenbush_hierarchical_2002,snijders_multilevel_2011}, can enhance DIF detection in CAT.
\\
DIF detection matters in two common CAT scenarios. First, the presence of item parameter drift \parencite[IPD;][]{bock_item_1988} requires ongoing monitoring and recalibration of CAT item pools to ensure fairness. This includes periodic reevaluation of DIF to maintain a DIF‑free pool \parencite{veerkamp_detection_2000}. Second, modern CAT programs increasingly use online calibration to field‑test new items during operational administrations \parencite{stocking_scale_1988,van_der_linden_shadow-test_2020}. In these settings, on‑the‑fly DIF detection becomes essential for evaluating item functioning during live testing. Benefits of online calibration, such as reduced sample‑size demands \parencite{ren_continuous_2017-2,vander_linden_optimal_2015} and increased operational efficiency \parencite{van_der_linden_shadow-test_2020}, would apply to real‑time DIF evaluation.
\\
The literature on DIF detection in CAT remains limited and largely rooted in early methodological adaptations of DIF procedures for linear tests. Most prior attempts relied on traditional or modified single‑level methods, yet these approaches have consistently demonstrated suboptimal performance in adaptive settings. For example, \textcite{zwick_simulation_1994} adapted the MH procedure using expected true scores from the full item pool---an approach that is impractical in CAT because examinees respond to only a small, adaptively selected subset of items. CATSIB \parencite{nandakumar_catsib_2001} showed initial promise, but its evaluation focused primarily on DIF effect size rather than item‑parameter differences. Subsequent studies \parencite{chu_detecting_2012,lei_comparing_2006} documented inflated Type‑I error rates for CATSIB as well as for LR and IRT‑LRT adaptations, highlighting unresolved challenges when applying single‑level DIF methods to CAT data. Despite substantial advances in DIF methodology more broadly, these limitations specific to DIF for CAT have not been fully addressed. A shared foundational limitation of the existing approaches is that they treat CAT item responses as independent, even though provisional ability estimates link items together adaptively, creating systematic dependencies within and across examinees. 
\\
In this article, we address this literature limitation by proposing a two-level logistic regression model to improve DIF detection in CAT by explicitly accounting for nuisance effects arising from CAT-induced structural dependency. We focus on DIF detection for pretested items in a unidimensional, item‑adaptive CAT environment. First, we conceptualize how between-item dependency associated with provisional ability estimates can be reframed as between-examinee dependency, and how multilevel modeling can be used to control for the resulting nuisance effects from this dependency. Then, we present a numeric example to illustrate how commonly used single‑level modeling can yield misleading inferences in CAT data---such as spurious DIF detection---as this dependency is ignored, and how our proposed two‑level modeling may mitigate this issue. Subsequently, we attempt Monte Carlo simulations to seek generalizable findings. We simulated the CAT data using {\fontfamily{qcr}\selectfont Pearson CAT 3.14} \parencite{ren_pearson_2017}, estimated single-level models using iteratively reweighted least squares (IWLS) via {\fontfamily{qcr}\selectfont glm} \parencite{r_core_team_r_2021}, and evaluated multilevel models via the Laplace approximation \parencite{raudenbush_maximum_2000} 
through {\fontfamily{qcr}\selectfont lme4} \parencite{bates_linear_2022-1} in R 4.2.1 \parencite{r_core_team_r_2021}. This work is expected to improve DIF detection in practical settings for CAT-driven large-scale assessments that have operational constraints (e.g., content balancing, exposure control).

\section{Dependencies in CAT}

In CAT, the selection of each item depends---partly or fully (given the specific item selection algorithm)---on the examinee’s previous responses through the provisional ability estimate, regardless of the purpose of the CAT algorithm (e.g., to optimize estimation precision, content balance, or exposure control). Before administering the $K$th item, the CAT algorithm computes a provisional estimate based on the first $(K-1)$ responses, and this estimate guides the next item choice. Because each provisional estimate is itself a function of earlier responses, items administered to the same examinee are \textit{statistically dependent} rather than conditionally independent. This sequential updating process creates \textit{between‑item dependency}: Information from earlier items influences the probability that particular items will be administered later in the test. As \textcite{mislevy_does_2000} already demonstrated, CAT data violate local independence when analyzed under a frequentist framework without accounting for the ordered, adaptive selection of items. 

\begin{figure}[htbp]
    \caption{Item Responses and Ability Estimates in a CAT}
    \label{fig:cat-item2}
    \includegraphics[width=0.5\textwidth]{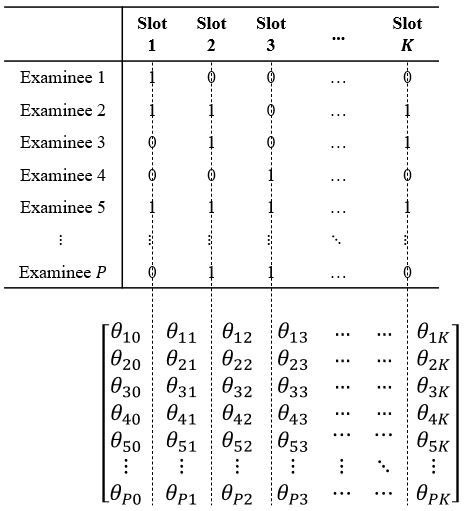}
\end{figure}

\begin{figure}[htbp]
\caption{A Range of Provisional Ability Estimates for a CAT Item}
    \label{fig:cat-item3}
    \includegraphics[width=\textwidth]{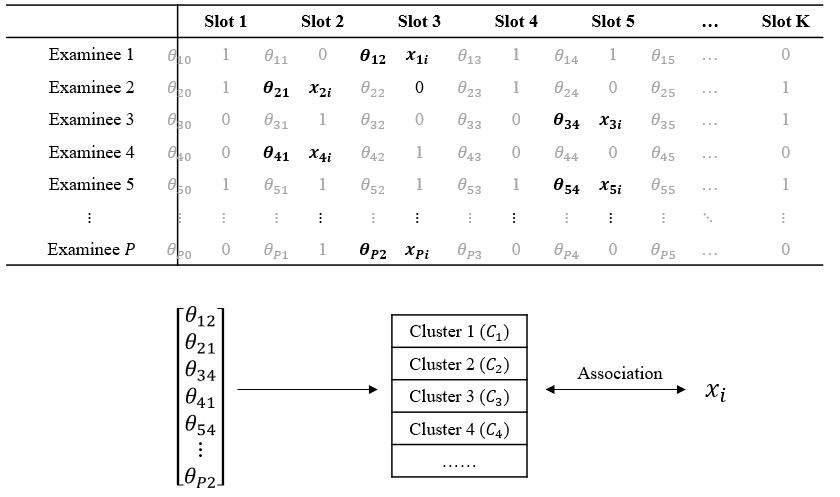}
\end{figure}

The structure becomes clearer when represented in matrix form, as shown in Figure \ref{fig:cat-item2}. Each row corresponds to an examinee and each column to an item slot, with the provisional ability estimate $\theta$ updating after each response. All examinees begin with the same starting estimate, $\theta_0$, as is standard in practice \parencite{van_der_linden_item_2000}. The provisional estimate $\theta_{ps}$ (where $s$ = 1, 2, $\cdots$, $K-1$) is updated iteratively after each administered item. The final ability estimate $\theta_{pK}$ is computed after the last administered item.
Item $i$ can appear at multiple slots for different examinees, as shown in Figure \ref{fig:cat-item3}. For example, Examinee 1 may receive item $i$ after Slot 2, associated with provisional estimate $\theta_{12}$; Examinee 2 may receive it at Slot 1 with $\theta_{21}$; other examinees may obtain this item at Slot 4 or 5, corresponding to $\theta_{34}$, $\theta_{41}$, $\theta_{54}$, and so on.
\\
To allow alternative DIF modeling without major changes to the adaptive algorithm or the related data structure, we see it useful to \textit{translate} between-item dependency for each examinee into \textit{between-examinee dependency} for the studied item. Specifically, we recognize that examinees tend to pass through similar ranges of provisional ability values when receiving a given item $i$. Examinees who encounter the same item at similar provisional ability estimates that immediately precede the item's administration, regardless of the item’s position in the test, can therefore be considered members of the same latent profile, proxied by the same provisional ability interval or cluster, denoted $C_j$ (where $j$ = 1, 2, 3, $\cdots, J$). In our proposed modeling approach, item responses $x_{pi}$, central to DIF analysis, should be analyzed in conjunction with provisional ability intervals, which function as \textit{macro‑level units} that capture the dependencies matched to each item. A given item may be administered across many such intervals, resulting in a nontrivial structural dependency. This is especially true when CAT algorithms incorporate operational constraints such as content balancing or exposure control, leading provisional ability estimates to vary substantially across examinees.

\section{A Two‑Level Logistic Model for DIF Detection in CAT}

The between‑item or between‑examinee dependencies results in a nested data structure, which has important implications for DIF modeling with CAT data, following \textcite{snijders_multilevel_2011}: If the influence of the macro‑level units is ignored, the resulting DIF estimates may be distorted by unmodeled cluster‑level variation. Specifically, the provisional‑ability interval $C_j$ may influence micro-level relations concerning DIF in multiple ways, known collectively as \textit{macro-to-micro interactions}. Traditional single‑level DIF methods ignore this nesting structure, as they model strictly micro-level associations among group membership, ability, and item response. They assume that item responses are independent once conditioning on total scores or final ability estimates, which are often suboptimal for matching examinees at the moment an item is administered. 
In effect, ignoring the nuisance effects via traditional single-level DIF models inflates error rates \parencite{dorman_effect_2008}, which underscores the need for a multilevel modeling approach to account for dependency induced by provisional‑ability estimates in CAT. 
\\
We propose a two-level framework for DIF analysis with CAT item response data. Multilevel modeling is particularly appropriate when fundamental assumptions required for single‑level models (e.g., independence of observations, homoscedasticity) are violated \parencite{raudenbush_hierarchical_2002}. These violations commonly occur in CAT, whereas traditional DIF methods such as MH, LR, SIBTEST, IRT-LR, or their modified versions \parencite[e.g.,][]{zwick_simulation_1994,nandakumar_catsib_2001,lei_comparing_2006} treat CAT item responses as independent. In contrast, multilevel models accommodate correlated errors from nested structures, do not require equal sample sizes or variances across higher‑level units \parencite{snijders_multilevel_2011}, allow the modeling of the influence of macro‑level units on micro‑level outcomes \parencite{gelman_data_2007}, and improve the generalizability of results as random effects are explicitly modeled \parencite{shadish_experimental_2002}. 

\subsection{Model Structure}

In our proposed two-level model, we estimate the fixed-effect DIF‑related parameters while accounting for the random effects associated with dependency among CAT item responses. The \textit{Level-2 units} are the provisional‑ability intervals associated with the studied item. These intervals represent the macro‑level contexts in which examinees encountered the item, and thus capture the between‑examinee dependency arising from CAT’s item‑selection process. Although discretizing a continuous provisional‑ability estimate inevitably sacrifices information, doing so produces a categorical clustering variable necessary for estimating random effects in multilevel modeling.
\\
Our \textit{Level-1 model} draws on variables commonly used in traditional DIF methods, namely group membership, ability, and group‑by‑ability interaction, from MH \parencite{holland_alternate_1985} and LR \parencite{swaminathan_detecting_1990}. However, we omit the final ability estimate from Level 1 due to its strong collinearity with provisional‑ability intervals, and we exclude the group‑by‑ability interaction because it reduces power for detecting uniform DIF \parencite{swaminathan_detecting_1990}. We retain the group membership variable $g$, which quantifies uniform DIF, and include a covariate for the interval size $n_j$ following \textcite[p. 56]{snijders_multilevel_2011}.
\\
Hence, our two-level model is expressed as follows:
\begin{equation}
\label{eqn:cat-dif}
    \begin{array}{lrll}
        \textbf{Level 1:} & & &\\
         & \mbox{logit}(\pi_{ij}) &=& \beta_{0j} + \beta_{1j} g + \beta_{2j} n_j \\
        \textbf{Level 2:} & & &\\
         & \beta_{0j} &=& \delta_{00}+U_{0j}  \\
         & \beta_{1j} &=& \delta_{10}+U_{1j}  \\
         & \beta_{2j} &=& \delta_{20} \\
    \end{array}
\end{equation}
where \\
\indent $g$ is the group membership, coded as 1 for the focal group and 0 for the reference group;\\
\indent $n_j$ is the cluster size for provisional‑ability interval $C_j$;\\
\indent $\delta_{00}$ is the grand mean log-odds of a correct response;\\
\indent $\delta_{10}$ is the average group effect (uniform DIF), associated with being in the focal group;\\
\indent $\delta_{20}$ is the effect of the cluster size; and\\
\hangindent=0.7in $U_{0j} \sim N(0, \tau_0^2)$ and $U_{1j} \sim N(0, \tau_1^2)$ are random intercept and random slope components.\\
\noindent The random effects here map onto two nuisance effects: (1) the macro‑level influence on the micro-level item response, and (2) the macro‑level moderation of the micro-level group effect (i.e., DIF).

\subsection{When Multilevel Modeling is Needed}

Multilevel modeling would be optimal to items with substantial between‑examinee dependency. The magnitude of between‑interval dependency can vary considerably across CAT items, as implied before when describing the range of provisional ability estimates. Items that appear frequently or at highly variable item‑slots are more vulnerable to dependency \parencite{chang_-stratified_1999}. This variability may explain why single‑level methods in \textcite{lei_comparing_2006} performed well for some items but not others.
\\
To determine whether a multilevel model is warranted, we propose the use of the intraclass correlation coefficient (ICC) or variance partition coefficient \parencite{goldstein_partitioning_2002} from an unconditional model
$\mbox{logit}(\pi_j)=\gamma_0+U_{0j}$,
following the recommendation from \textcite{snijders_multilevel_2011}.
The ICC for two‑level logistic regression is
\begin{equation}
    \rho(y) = \dfrac{\tau^2}{\tau^2 + \pi^2/3},
\end{equation}
where $\tau^2$ is the between‑interval variance and $\pi^2/3$ is the logistic variance \parencite{goldstein_partitioning_2002}.
ICC values above roughly .10 are often considered substantively meaningful in educational research, where values between .10 and .25 are commonly observed \parencite{hedges_intraclass_2007}. 
In this study, they indicate substantial clustering that can inflate standard errors and bias DIF estimates.

\section{A Numeric Example}

Below we present a numeric example based on a simulated CAT dataset designed to reflect a practical CAT setting and intentionally constructed to be free of DIF. By comparing single‑level and multilevel models applied to this dataset, we demonstrate how modeling examinees as nested within provisional‑ability intervals provides a principled way to account for structural dependency in CAT data and mitigate resulting nuisance effects.

\subsection{CAT Structure and Data Generation}

To ensure realistic CAT behavior, our item selection was governed by content‑balanced and exposure‑controlled algorithms commonly used in operational settings. We used the weighted penalty model \parencite[WPM;][]{shin_weighted_2009} that involves content constraints in item selection and a conditional randomesque strategy \parencite{shin_comparison_2012,shin_conditional_2017} to manage item exposure. The content and information indices in the WPM criterion were equally weighted, resulting in a moderately constrained item selection mechanism. Content constraints from the TIMSS 2019 test specification for Grade 8 were considered. They spanned four content domains and three cognitive domains, with each domain comprising approximately 20\%–40\% of the item pool \parencite{mullis_timss_2017}.  
The CAT item pool consisted of 200 items drawn from the TIMSS 2019 Grade-8 Mathematics assessment \parencite{fishbein_timss_2021}, including 189 binary items and 11 polytomous items. Once administered, items were excluded from subsequent selection. Following standard CAT practice, all examinees began with the same initial ability estimate $\theta_0=0$ \parencite{van_der_linden_item_2000}. True abilities were sampled from a standard normal distribution $N (0, 1)$. The provisional and final ability values ranged approximately from -4 to 4, estimated via MLE. The maximum item exposure rate was set to be .33.
\\
We simulated CAT item responses from 5,000 examinees, each administered a fixed‑length 25‑item CAT. The resulting data exhibited desirable properties, with an overall ability‑estimation bias of .04, a mean squared error (MSE) of .17, a correlation of .93 between true and estimated final ability, and an average conditional standard error of measurement (CSEM) of .37. These values are consistent with the performance levels typically reported for the WPM and competing CAT item‑selection algorithms \parencite[e.g.,][]{shin_weighted_2009}. Given this data, we produced provisional‑ability intervals using 81 equally-spaced points with a width of .1 for the ability range [-4, 4].
The focal group was defined as females, with examinees randomly assigned such that half of the examinees were male and half were female.

\subsection{Data Cleaning}

\begin{figure}[ht!]
\caption{\label{fig:cat-iccs}Distribution of ICCs across Retained Items}
    \includegraphics[width=0.8\textwidth]{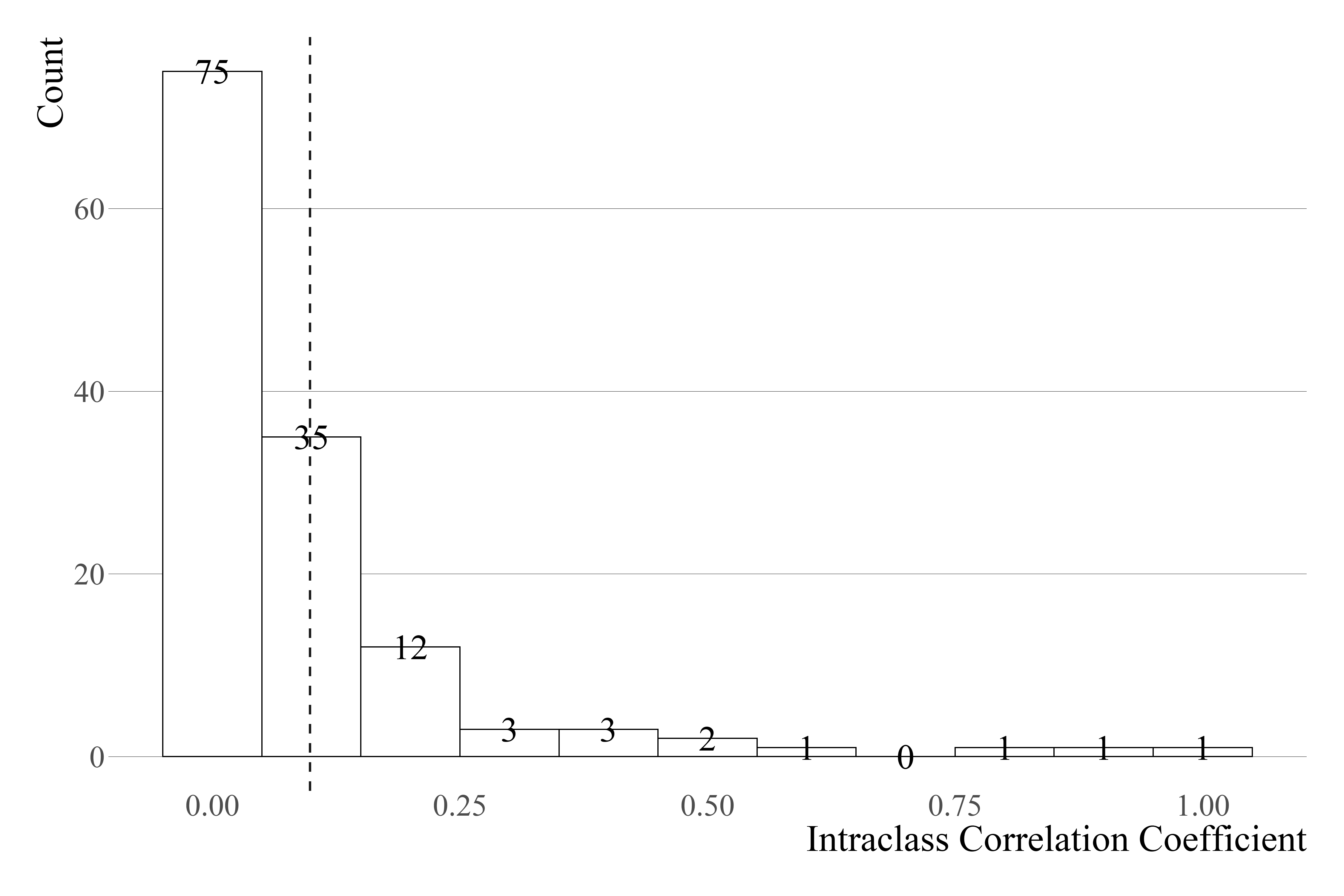}
    \text{\textit{Note}. The bin width is .1. The blue dashed line denotes the mean of $\rho(y)$ values.}
\end{figure}

During data cleaning, we took the following steps to keep items that would potentially benefit from multilevel modeling for illustrative purposes: We removed polytomous items, the first administered item for each examinee (which does not depend on provisional ability), items with insufficient variance or too few provisional‑ability intervals (i.e., a single level‑2 unit or a single response category), and items with too few observations to support model estimation. In addition, we retained only those administered to at least 1,000 examinees and associated with at least 50 unique level‑2 units after CAT administration, as fixed‑effect parameter estimates in multilevel logistic regression are reliable under these conditions 
In the end, we retained 134 items 
\parencite{moineddin_simulation_2007,ali_sample_2019}.
\\
High‑ICC items are likely to produce biased DIF estimates if analyzed using single‑level models, and thus would serve as ideal examples for applying the proposed multilevel framework. Hence, we computed ICC for each of these retained items using the unconditional model described before. The distribution of ICC values was found to be positively skewed, as shown in Figure \ref{fig:cat-iccs}. Approximately 13\% exhibited ICCs greater than .20, a level typically considered meaningful in educational measurement contexts \parencite{hedges_intraclass_2007}.

\subsection{Descriptive Statistics}

\begin{figure}[htbp]
    \caption{Distribution of Examinees across Provisional Ability Intervals}
    \label{fig:cat-barplot}
    \includegraphics[width=0.8\textwidth]{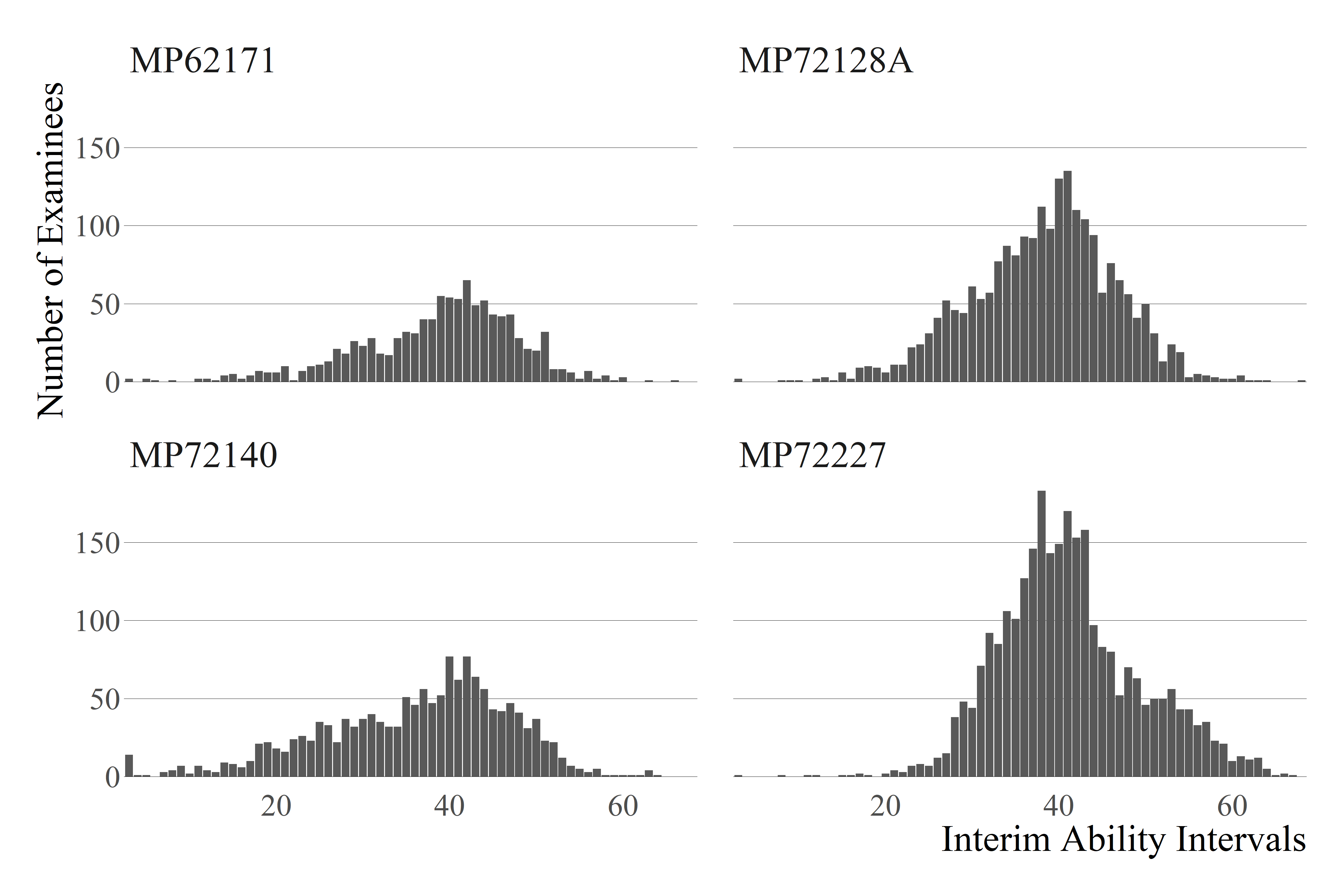}
    
    \text{\textit{Note}. Provisional ability intervals were defined using 81 equally spaced points}
    \text{(step size = .1) spanning [-4, 4].}
\end{figure}

Only a small subset of representative items is discussed in this numeric example for a detailed evaluation. Specifically, four items that met stricter criteria ($\geqslant$ 50 provisional ability intervals, $\geqslant$ 1,000 examinees) are presented in Figure \ref{fig:cat-barplot}, a barplot showing the number of examinees per provisional ability interval for these items. The $y$-axis represents the number of examinees, and the $x$-axis displays the provisional ability intervals ordered from low to high across the 81 possible intervals, noting that each item occupies a subset of these intervals. Many intervals in the middle of the $x$-axis had more than 50 examinees, which is an ideal cluster size \parencite{moineddin_simulation_2007,ali_sample_2019}. While the item‑specific distributions were somewhat skewed, they were generally unimodal with the largest counts occurring near the middle, which roughly aligned with the expectation that random effects at the interval level are normally distributed.

\begin{sidewaysfigure}[htbp]
    \caption{Item Characteristic Curves of Item MP72227 for Two Groups (Male vs. Female)}
    \label{fig:cat-itemchar}
    \includegraphics[width=\textwidth]{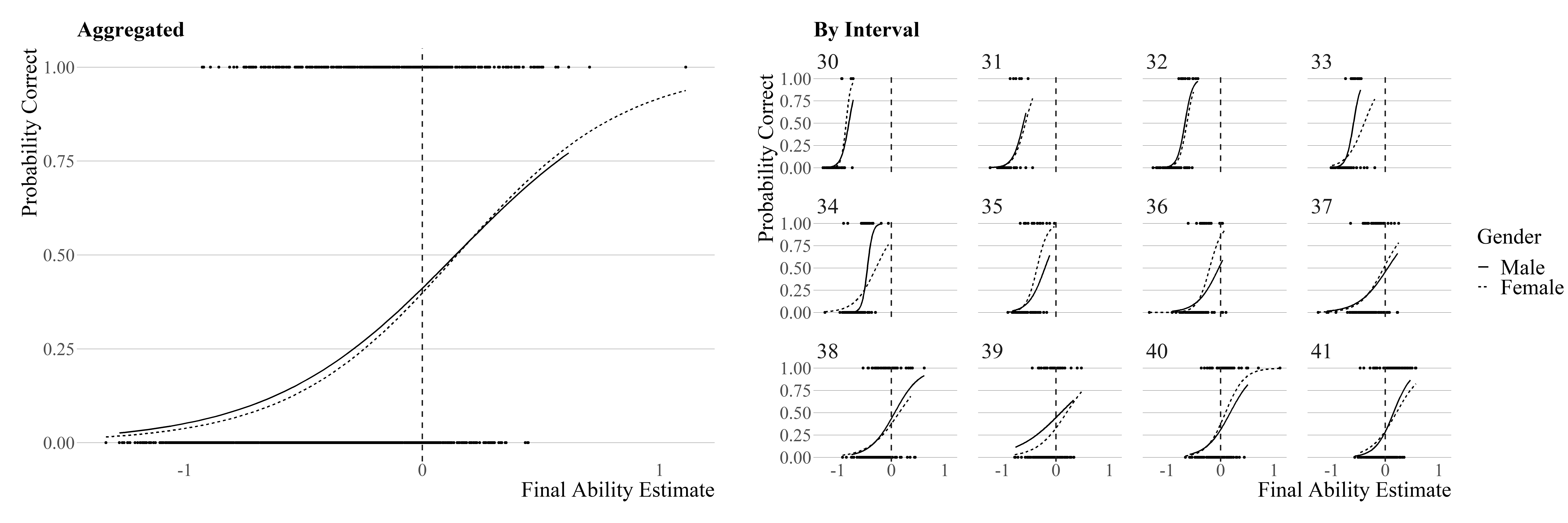}
    \textit{Note}. The number at the top of each curve plot indicates the rank of the provisional ability interval. The blue dashed vertical line at $x$ = 0 serves as a reference.
\end{sidewaysfigure}

Item characteristic curves for the item with the highest $\rho(y)$, MP72227, are displayed in Figure \ref{fig:cat-itemchar} for illustrative purposes. The left panel shows curves based on data from all examinees, whereas the right panel presents curves for examinees in the 30th to 41st provisional ability intervals. Not all intervals are shown due to space constraints. The $x$-axis represents examinees’ final ability estimate, and the $y$-axis shows the proportion of examinees who answered the item correctly. We observed clear differences between males and females in the probability of a correct response when holding the final ability estimate constant. Meanwhile, both the magnitude (larger vs. smaller differences) and the nature (intersecting vs. non-intersecting curves) of these differences varied across provisional ability intervals. This interval-specific variation suggests that DIF estimate may differ depending on provisional ability intervals in which the item is administered, highlighting the value of multilevel modeling for obtaining more accurate DIF estimates by accounting for interval-specific nuisance effects.

\subsection{Modeling}

We estimated a series of models to evaluate the appropriateness of our proposed two-level model in Equation (\ref{eqn:cat-dif}), using data from item MP72140 for illustration. To provide a principled basis for comparison, we adopted a multimodel inference perspective in which no single specification is assumed a priori to be the true data-generating model \parencite{burham_david_2004}. Accordingly, models were compared using a combination of (a) convergence (only converged solutions were considered interpretable), (b) model fit statistics, and (c) substantive decision rules aligned with the known data‑generating condition (i.e., the dataset was simulated without DIF), with preference given to models that avoid spurious DIF detection.
\\
Single-level modeling results are in Table \ref{tab:single}. Model S1 was adapted from the logistic regression DIF model proposed by \textcite{swaminathan_detecting_1990}, using the final CAT ability estimate $\theta_K$ as a proxy for latent ability \parencite{lei_comparing_2006}. Model S2 incorporated the provisional ability $\theta_s$ as a continuous covariate to test whether controlling for provisional ability at the micro level would improve DIF estimation. Model S3 was similar to Model S1 but replaced the final ability $\theta_K$ with the provisional ability $\theta_s$. We showed that Models S1 and S3 yielded statistically significant coefficients for the group indicator $g$, suggesting uniform DIF, which were false positives because the simulated data did not contain DIF. Although the coefficient for $g$ in Model S2 was not statistically significant, it was close to the .05 threshold, indicating a substantial risk of reaching false positives when including provisional ability as a micro-level covariate instead of macro-level units.
\begin{table}[htbp]
\small
\centering
\begin{threeparttable}
\caption{\label{tab:single}Single-Level Modeling Results}
\def\arraystretch{0.6} 
\begin{tabular}{llll}
\toprule
 & \multicolumn{1}{c}{S1} & \multicolumn{1}{c}{S2} & \multicolumn{1}{c}{S3} \\
\midrule
 (Intercept) & -0.362$^{***}$ & -0.487$^{***}$ & -0.395$^{***}$ \\ 
  & (0.081) & (0.088) & (0.081) \\ [1.8ex] 

 $g$ & 0.261$^{*}$ & 0.221$^{.}$ & 0.260$^{**}$ \\ 
  & (0.117) & (0.126) & (0.116) \\ [1.8ex] 
  
 $\theta_K$ & 0.644$^{***}$ & 10.121$^{***}$ &  \\ 
  & (0.085) & (0.789) &  \\ [1.8ex] 

 $\theta_K g$ & 0.245$^{.}$ & 0.257$^{.}$ &  \\ 
  & (0.127) & (0.156) &  \\ [1.8ex] 

 $\theta_s$  & & -9.056$^{***}$ & 0.494$^{***}$ \\ 
  &  & (0.748) & (0.079) \\ [1.8ex] 

 $\theta_s g$ &  &  & 0.247$^{*}$ \\ 
  &  &  & (0.118) \\ [1.8ex] 

\midrule
AIC & 1,785.35 & 1,527.79 & 1,841.99  \\  
BIC & 1,806.55 & 1,554.29 & 1,863.19 \\
Deviance  &  1,777.35 & 1,517.79 & 1,833.99  \\
N. of Units & 1,480 & 1,480 & 1,480 \\
\bottomrule
\end{tabular}
    \begin{tablenotes}
    \vspace{4pt}
    \text{\textit{Note}. $^{.}p<0.1$; $^{*}p<0.05$; $^{**}p<0.01$; $^{***}p<0.001$.}
    \end{tablenotes}
\end{threeparttable}  
\end{table}

Multilevel modeling results are in Table \ref{tab:multi}. Models M1--M4 included random intercepts only, whereas Models M5–M8 included random slopes for the group indicator $g$. Models M3--M4 and M7--M8 additionally controlled for cluster size. Notably, several models that included the final ability estimate $\theta_K$ produced statistically significant coefficients for $g$, incorrectly indicating DIF. This result was likely attributable to strong collinearity between $\theta_K$ and the provisional ability intervals defining the level-2 structure, which can inflate standard errors and bias fixed-effect estimates. Model M6---our proposed two-level model in Equation (\ref{eqn:cat-dif})---demonstrated the strongest performance. Its coefficient for $g$ was not statistically significant, correctly reflecting the absence of DIF in the data. Compared to Model M5, which also yielded a non-significant group effect, M6 is theoretically preferable because it controls for cluster size in addition to modeling random intercepts and random slopes corresponding to nuisance effects described earlier. Among the models not affected by collinearity (M1--M2 and M5--M6), M6 exhibited the smallest intercept variance ($\tau_0^2$ = 0.277) and a relatively high conditional $R^2$ value ($R_{cond}^2$ = .146), indicating superior explanatory power. 
These results provided strong empirical support for our proposed two-level model. Consistent with this finding, a study by \textcite{chen_modeling_2023} showed that Model M6 outperformed CATSIB when applied to the same simulated data set.
\begin{landscape}
\begin{table}[htbp]
\small
\def\arraystretch{0.6} 
\begin{threeparttable}
\caption{\label{tab:multi}Multilevel Modeling Results}
\begin{tabular}{lllllllll}
\toprule
 & \multicolumn{1}{c}{M1} & \multicolumn{1}{c}{M2} & \multicolumn{1}{c}{M3} & \multicolumn{1}{c}{M4} & \multicolumn{1}{c}{M5} & \multicolumn{1}{c}{M6} & \multicolumn{1}{c}{M7} & \multicolumn{1}{c}{M8} \\
\midrule

 (Intercept) & -0.707$^{***}$ & -0.971$^{***}$ & -0.360$^{***}$ & 0.004 & -0.660$^{***}$ & -0.950$^{***}$ & -0.366$^{***}$ & -0.022 \\ 
  & (0.135) & (0.241) & (0.084) & (0.173) & (0.119) & (0.233) & (0.083) & (0.180) \\ [1.8ex] 

 $g$ & 0.205$^{.}$ & 0.201$^{.}$ & 0.264$^{*}$ & 0.287$^{*}$ & 0.135 & 0.163 & 0.314$^{*}$ & 0.312$^{*}$ \\ 
  & (0.113) & (0.113) & (0.118) & (0.119) & (0.134) & (0.132) & (0.141) & (0.138) \\ [1.8ex] 

 $n_j$ &  & 0.008 &  & -0.008$^{*}$ &  & 0.008 &  & -0.008$^{*}$ \\ 
  &  & (0.006) &  & (0.003) &  & (0.006) &  & (0.004) \\ [1.8ex] 

 $\theta_K$ &  &  & 0.647$^{***}$ & 0.682$^{***}$ &  &  & 0.642$^{***}$ & 0.677$^{***}$ \\ 
  &  &  & (0.086) & (0.085) &  &  & (0.085) & (0.086) \\ [1.8ex] 

 $\theta_K g$ &  &  & 0.247$^{.}$ & 0.267$^{*}$ &  &  & 0.294$^{*}$ & 0.295$^{*}$ \\ 
  &  &  & (0.127) & (0.126) &  &  & (0.142) & (0.139) \\ [1.8ex] 
  
  \midrule
AIC & 1,918.55 & 1,918.83 & 1,787.31 & 1,783.44 & 1,917.99 & 1,917.93 & 1,788.46 & 1,785.86 \\ 
BIC & 1,934.45 & 1,940.03 & 1,813.81 & 1,815.24 & 1,944.49 & 1,949.72 & 1,825.56 & 1,828.26  \\ 
Deviance & 1,912.54 & 1,910.82 & 1,777.3 & 1,771.44 & 1,907.98 & 1,905.92 & 1,774.46 & 1,769.86 \\
N. of Level-1 Units & 1,480 & 1,480 & 1,480 & 1,480 & 1,480 & 1,480 & 1,480 & 1,480 \\ 
N. of Level-2 Units &  61 & 61 & 61 & 61 & 61 & 61 & 61 & 61 \\

\midrule
$\tau_0^2$  & 0.514 & 0.473 & 0.005 & 0.000 & 0.319 & 0.277 & 0.001 & 0.006 \\
$\tau_1^2$   &  &  &  &  & 0.15 & 0.152 & 0.123 & 0.111 \\
$\tau_{10}$ &  &  &  &  & 0.17 & 0.177 & -0.013 & -0.026 \\
ICC & .135 & .126 & .002 &  --- & .148 & .139 & .015 & .011 \\
$R^2_{marg}$ & .003 & .009 & .187 & .186 & .001 & .008 & .194 & .190 \\ 
$R^2_{cond}$ & .138 & .134 & .188 &  --- & .149 & .146 & .206 & .199\\ 
\bottomrule
\end{tabular}
    \begin{tablenotes}
    \vspace{4pt}
     \text{\textit{Note}. $^{.}p<.1$; $^{*}p<.05$; $^{**}p<.01$; $^{***}p<.001$.}
    \end{tablenotes}
\end{threeparttable}  
\end{table}
\end{landscape}

\section{Monte Carlo Simulation}

The numeric example was intentionally designed to be DIF‑free and tied to a specific CAT configuration (e.g., fixed test length, particular exposure rate and ability estimator). It also relied on a restricted subset of items that satisfied feasibility requirements for multilevel estimation (e.g., sufficient provisional‑ability intervals and sample size), with model comparisons illustrated using a single representative item. These features make the example well-suited for demonstrating how CAT‑induced dependency can lead to misleading DIF inferences under common single‑level models, but they limit generalizability to broader operational CAT settings in which test length, exposure rate, ability estimator, and DIF contamination vary and cluster structures may be uneven.
\\
In seeking generalizable patterns, we conducted Monte Carlo simulations that systematically varied key CAT and DIF factors (i.e., test length, exposure control, ability estimator, DIF type, and DIF prevalence), employed a substantially larger item pool, and evaluated performance in terms of item-level Type‑I error and statistical power computed across replications and averaged across items. Each simulated condition was replicated 100 times\footnote{Each condition was replicated 100 times, a replication count that is commonly used in DIF simulation studies evaluating Type-I error and power \parencite[e.g.,][]{lim_2024}. Also, for a binomial detection outcome, the Monte Carlo standard error of an estimated Type-I error rate is $SE(\hat{p}) = \sqrt{p(1-p)/R}$; at $p = .05$ and $R = 100$, $SE(\hat{p}) \approx .022$, which is small enough to provide adequate precision for comparing methods. In addition, because we summarize item-level Type-I error and power by averaging across eligible items within each condition, the resulting estimates are further stabilized.} with 5,000 examinees. We compared the best‑performing multilevel model from the numeric example, M6, with three competing single‑level models: (1) S1, an LR model \parencite{swaminathan_detecting_1990} using the final ability estimate as the matching criterion; (2) S2, which incorporates provisional ability as a covariate; and (3) S3, an LR model using provisional ability as the matching criterion. 
\\
For comparability across models, item‑level Type‑I error rates and statistical powers were computed using only those replications in which all candidate models converged. As a result, the effective item count (in both Study 1 and Study 2) and replication count (in Study 2) varied by condition, reflecting differences in model convergence behavior across various CAT data structures. We did not attempt to replace nonconvergent replications with additional simulated runs, as we treated convergence as an integral component of practical applicability and reported results conditional on joint convergence, consistent with how DIF models would be applied in practice.

\subsection{Simulation Design}

\textbf{Study 1: Evaluation of Type-I error rate.} Study 1 was designed to evaluate the Type‑I error control of the proposed two‑level model (M6) relative to competing single‑level approaches under DIF‑free conditions. Type‑I error was defined as the proportion of DIF‑free items incorrectly flagged as exhibiting DIF, based on a statistically significant group effect at the .05 level. Item-level Type‑I error rates were computed for each DIF‑free item across replications and summarized using the mean and standard deviation across eligible items within each condition.
\\
Drawing from common discussions in the CAT and DIF literature \parencite[e.g.,][]{van_der_linden_assembling_2006,svetina_detecting_2014,woods_langer-improved_2013}, we manipulated the following factors in Study 1: (1) provisional ability estimation method (MLE vs. EAP), 
(2) CAT test length (25 vs. 35 items), and (3) maximum exposure rate (.20 vs. .33). 
This design resulted in eight simulated conditions (2 $\times$ 2 $\times$ 2). These factors were selected because they represent core operational CAT design decisions that are expected to influence parameter estimation and, in turn, DIF detection.
\\
\textit{Test length (25 vs.\ 35 items).} Test length was manipulated because longer tests generally yield more stable ability estimation and reduced bias under sufficiently large item pools \parencite{chang_nonlinear_2009,chang_psychometrics_2015}, which can affect both the magnitude of adaptivity-induced dependency and the reliability of DIF estimation. Shorter tests also place greater weight on early-item information when ability estimates are less precise, potentially amplifying model misspecification effects.
\\
\textit{Maximum exposure rate (.20 vs.\ .33).} Exposure control was manipulated because higher allowable exposure increases item reuse, which can intensify dependence patterns induced by adaptive item selection and provisional scoring. In contrast, lower exposure limits can diversify administered items but may increase heterogeneity in item paths and cluster structures. The chosen values (.20 and .33) reflect commonly used exposure constraints in CAT settings.
\\
\textit{Ability estimation method (MLE vs.\ EAP).} Provisional ability $\theta_s$ was estimated using either MLE or EAP because the ability estimator can affect between-item dependency in CAT and the degree to which local independence is violated in practice \parencite{mislevy_does_2000}. We contrasted MLE, a frequentist approach commonly used in CAT simulations and operational scoring, with EAP, a Bayesian estimator frequently recommended for stable interim scoring and information-based selection \parencite[p. 364]{kingsbury_procedures_1989-2}.
\\
\textbf{Study 2: Evaluation of statistical power.} Study 2 extended the simulation design to evaluate statistical power under DIF‑contaminated conditions. Statistical power was defined as the proportion of DIF‑contaminated items correctly identified as exhibiting DIF, based on a statistically significant group effect at the .05 level. Power was calculated for each DIF‑contaminated item across replications and summarized using the mean and standard deviation across eligible items. In addition to the factors manipulated in Study 1, we introduced two DIF‑related factors in Study 2: (4) DIF type, represented by a between‑group difference of .4 in either Parameter $a$ (discrimination) or $b$ (difficulty) , and (5) DIF prevalence (20\% vs. 40\% of the administered items in one CAT administration). This design resulted in 32 simulated conditions (8 $\times$ 2 $\times$ 2). These added factors were selected to reflect practically relevant DIF conditions.
\\
\textit{DIF type (DIF in $a$ vs. $b$).} DIF occurrences in Parameters $a$ and $b$ represent substantively different forms of DIF and can differentially affect detectability under logistic and multilevel DIF models. Following \textcite{penfield_assessing_2001}, we implemented DIF through parameter differences rather than manipulating effect sizes; either Parameter $a$ or $b$ was inflated in the focal group to simulate DIF.
\\
\textit{DIF prevalence (20\% vs.\ 40\%).} DIF prevalence reflects variation in how widely DIF may be distributed and can affect detection. Higher DIF prevalence can alter item-selection behavior, which has direct implications for statistical power and model convergence in realistic CAT settings. DIF-contaminated CAT items were randomly selected in each replication.
\\
\textit{Difference of .4 in parameters.} DIF magnitude was operationalized as a between‑group difference of .4 in either Parameter $a$ or $b$. This value was chosen to represent a moderate level of DIF \parencite{lim_2024} that is substantively meaningful but not extreme, allowing evaluation of statistical power without guaranteeing detection. Parameter differences of this magnitude are commonly used in DIF simulation studies to assess method sensitivity under realistic conditions \parencite[e.g.,][]{penfield_assessing_2001}.

\subsection{CAT Structure, Data Generation, and Data Cleaning}

\begin{figure}[H]
    \caption{Distribution of Item Parameters in the 800-Item Pool}
    \label{fig:cat-itempar}
    \includegraphics[width=0.9\textwidth]{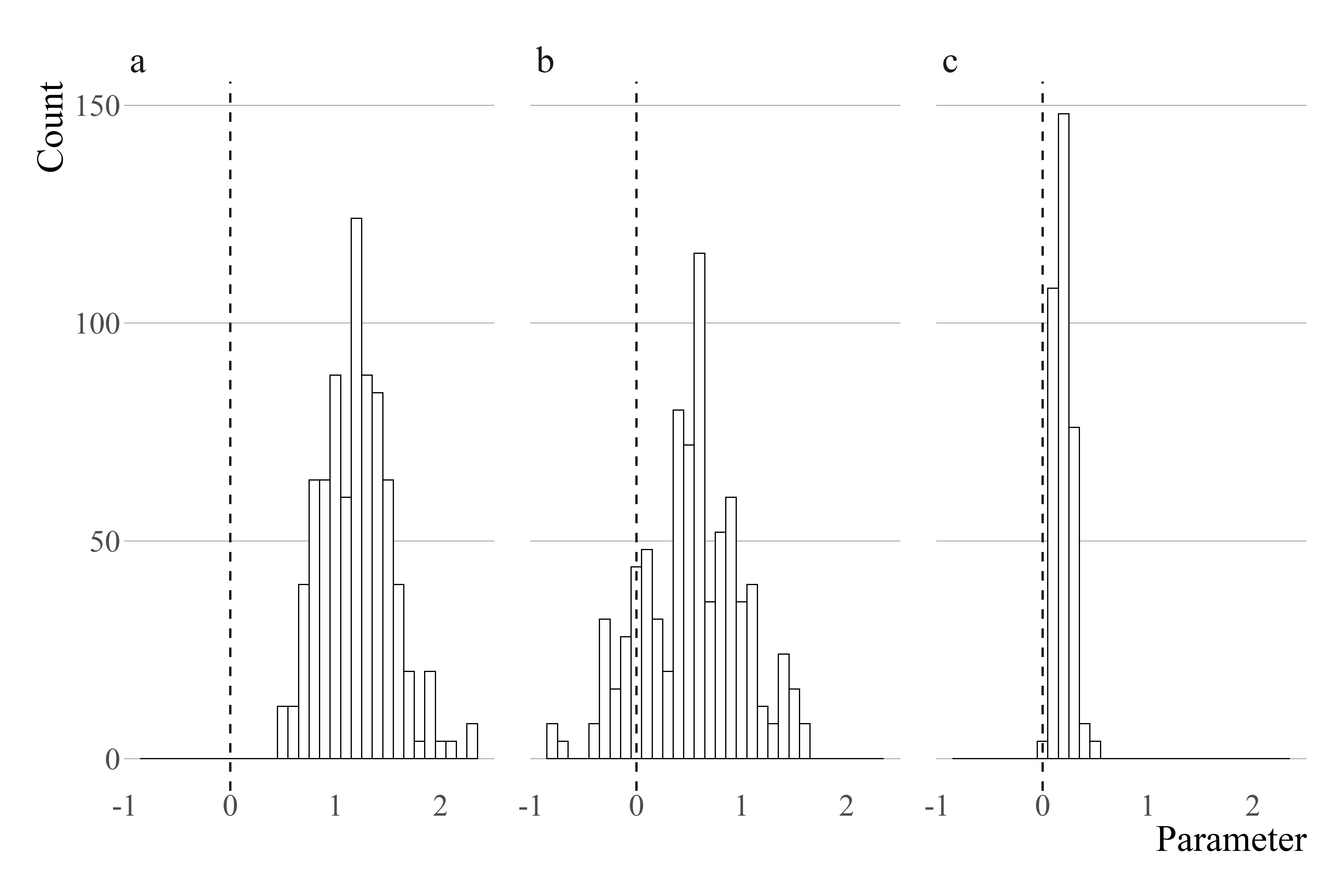}
\end{figure}

All CAT settings not explicitly manipulated in the simulation were held constant and matched those used in the numeric example. Specifically, content constraints were derived from the TIMSS 2019 test specifications; item selection was implemented using the WPM combined with a conditional randomesque strategy; all examinees began with the same initial ability value; true abilities were sampled from a standard normal distribution $N(0, 1)$ with values ranging approximately from -4 to 4; and final ability estimates were obtained using MLE.
\\
For the Monte Carlo simulations, we used an expanded item pool of 800 items, four times larger than the original 200‑item pool employed in the numeric example. The expanded pool was created by replicating the original set of 102 two‑parameter logistic (2PL) items, 87 three‑parameter logistic (3PL) items, and 11 generalized partial credit (GPC) items four times and relabeling them to form unique items. 
Expanding the pool was necessary to meet recommended CAT pool size ratios \parencite{stocking_three_1994}, support 25- and 35-item CAT tests \parencite{van_der_linden_assembling_2006}, and improve test quality \parencite{chang_comparative_2003}. 
Figure \ref{fig:cat-itempar} illustrates the empirical distribution of item parameters in the expanded pool, which are consistent with values commonly observed in large‑scale assessments.

For DIF-free conditions in Study 1, item responses were simulated as items were adaptively selected using the CAT engine. For DIF-contaminated conditions in Study 2, reverse engineering was applied to generate item responses based on DIF-affected item parameters. Specifically, the DIF-contaminated item parameters were treated as fixed inputs to the CAT engine, which was then used to generate item response data reflecting the impact of DIF on examinee responses, reversing the typical process of estimating item parameters from observed responses. The resulting CAT data were then prepared for multilevel modeling using the same data cleaning procedures as in the numeric example, with one exception: Items were not excluded based on minimum cluster size or examinee counts, allowing for preserving all the heterogeneous cluster structures observed in practice.

\subsection{Simulation Results}

\textbf{Quality of CAT data.} Before interpreting DIF results, we verified that the simulated CAT administrations produced adequate measurement precision under DIF-free conditions. Table \ref{tab:cat_stat} reports bias, MSE, correlations between estimated and true ability, and CSEM averaged across the 100 replications. Across all DIF-free conditions, bias remained small (.02--.03), MSE values were modest (.10--.14), correlations were high (.94--.96), and CSEM values were low (.28--.34), indicating that the CAT configurations yielded stable ability estimation consistent with prior CAT simulation evidence \parencite[e.g.,][]{shin_weighted_2009}. These results clarified that differences in DIF outcomes across models and conditions were not driven by poor CAT measurement precision in the baseline settings.

\begin{table}[htbp]
\def\arraystretch{0.6} 
\footnotesize
    \caption{\label{tab:cat_stat}Measurement Precision in CAT Data Simulation (Study 1, 100 Replications)}
\begin{tabular}{cccrrrr}
\hline
\multicolumn{1}{c}{Est.} & \multicolumn{1}{c}{Test} & \multicolumn{1}{c}{Max.} & \multicolumn{4}{c}{Precision} \\
\cmidrule(r){4-7}
Method & Length & Exp. Rate & Bias & MSE & Correlation & CSEM\\
\hline
MLE & 25 & .20 & .025 & .144 & .941 & .342\\

   &  &  & (.006) & (.005) & (.002) & (.001)\\
 
 &  &  .33 & .026 & .144 & .941 & .340\\

   &  &  & (.005) & (.005) & (.002) & (.001)\\

 & 35 & .20 & .017 & .095 & .958 & .283\\

 &  &  & (.004) & (.003) & (.001) & (.001)\\
 
 &  & .33 & .017 & .095 & .959 & .281\\

 &  &  & (.004) & (.003) & (.001) & (.001)\\

EAP & 25 & .20 & .025 & .142 & .941 & .341\\

 &  &  & (.005) & (.005) & (.002) & (.001)\\
 
 &  & .33 & .025 & .142 & .941 & .340\\

 &  &  & (.005) & (.005) & (.002) & (.002)\\

 & 35 & .20 & .017 & .094 & .959 & .282\\

 &  &  & (.004) & (.003) & (.001) & (.001)\\
 
 &   & .33 & .017 & .093 & .959 & .281\\

 &  &  & (.005) & (.003) & (.001) & (.001)\\

\hline
\end{tabular}
\end{table}

\textbf{Study 1.} Our primary conclusion from Study 1 is that, the proposed multilevel model (M6) more consistently controls false-positive rates than single-level models when no DIF is present. As shown in Table \ref{tab:chp4_nDIF_typ1}, across all the eight CAT conditions, all models produced Type-I error rates within a commonly acceptable range, but M6 was uniformly the most conservative (i.e., lowest mean Type-I error) across test length, exposure rate, and ability estimator combinations. This pattern is consistent with the central claim of this article: Explicitly modeling examinees as nested within provisional ability intervals mitigates nuisance dependency. Practically, we demonstrated that the multilevel adjustment is advantageous for protecting against over-flagging DIF-free items in CAT administrations.

\begin{table}[htbp]
\footnotesize
\def\arraystretch{0.6} 
\centering
\begin{threeparttable}  
\caption{\label{tab:chp4_nDIF_typ1}Item-Level Type-I Error Rate (Study 1, 100 Replications)}
\begin{tabular}{lllrrrrrrrrrrrr}
\hline
\multicolumn{1}{c}{ } & \multicolumn{1}{c}{ } & \multicolumn{1}{c}{ } & \multicolumn{12}{c}{Models} \\
\multicolumn{1}{c}{Est.} & \multicolumn{1}{c}{Test} & \multicolumn{1}{c}{Max.} & \multicolumn{3}{c}{M6} & \multicolumn{3}{c}{S1} & \multicolumn{3}{c}{S2} & \multicolumn{3}{c}{S3} \\
\cmidrule(r){4-6} \cmidrule(r){7-9} \cmidrule(r){10-12} \cmidrule(r){13-15}
Method & Length & Exp. & $\mu$ & $\sigma$ & $n$\tnote{i} & $\mu$ & $\sigma$ & $n$ & $\mu$ & $\sigma$ & $n$ & $\mu$ & $\sigma$ & $n$\\
\hline
MLE & 25 & .20 & .039 & .023 & 207 & .047 & .027 & 207 & .051 & .030 & 207 & .047 & .024 & 207\\
 
 &  & .33 & .042 & .026 & 171 & .050 & .029 & 171 & .054 & .028 & 171 & .048 & .023 & 171\\

 & 35  & .20 & .045 & .025 & 236 & .046 & .024 & 236 & .049 & .026 & 236 & .047 & .023 & 236\\
 
 &  & .33 & .043 & .026 & 206 & .048 & .024 & 206 & .049 & .028 & 206 & .046 & .022 & 206\\

EAP & 25 & .20 & .041 & .025 & 197 & .047 & .026 & 197 & .050 & .030 & 197 & .046 & .024 & 197\\
 
    &  & .33 & .043 & .031 & 156 & .049 & .026 & 156 & .053 & .032 & 156 & .046 & .022 & 156\\

& 35 & .20 & .044 & .024 & 224 & .050 & .028 & 224 & .054 & .030 & 224 & .047 & .024 & 224\\

 &  & .33 & .044 & .029 & 190 & .047 & .026 & 190 & .052 & .030 & 190 & .047 & .024 & 190\\

\hline
\end{tabular}
\begin{tablenotes}
\vspace{4pt}
        \textit{Note}. \textsuperscript{i} Effective item count across the 100 replications.
    \end{tablenotes}
\end{threeparttable}
\end{table}

\textbf{Study 2.} Across the Study 2 conditions, DIF detectability in CAT is jointly determined by various factors, as shown Tables \ref{tab:chp4_wDIF_typ1}--\ref{tab:chp4_wDIF_power}. M6 was advantageous when the test was short (25 items), with a small number of DIF items (20\%) and a low item exposure rate (.20). However, power was low for all the models, which we further discussed together with model convergence in the next section. In addition, when 20\% of items were DIF-contaminated, power was generally modest and declined as the maximum exposure rate increased from .20 to .33, particularly for shorter tests (25 items), reflecting that limited information from the test administration can negatively affect power. When DIF prevalence increased to 40\%, power deteriorated further across nearly all conditions and no model consistently performed better than others, suggesting that widespread contamination disrupted the adaptive process and reduced the stability of item-level DIF estimation. 
\begin{table}[htbp]
\footnotesize
\def\arraystretch{0.6} 
\centering
\begin{threeparttable}  
\caption{\label{tab:chp4_wDIF_typ1}Item-Level Type-I Error Rate (Study 2, 100 Replications)}
\begin{tabular}[t]{lllrrrrrrrrrrrr}
\hline
\multicolumn{1}{c}{ } & \multicolumn{1}{c}{ } & \multicolumn{1}{c}{ } & \multicolumn{12}{c}{Models} \\
\multicolumn{1}{c}{Est.} & \multicolumn{1}{c}{Test} & \multicolumn{1}{c}{Max.} & \multicolumn{3}{c}{M6} & \multicolumn{3}{c}{S1} & \multicolumn{3}{c}{S2} & \multicolumn{3}{c}{S3} \\
\cmidrule(r){4-6} \cmidrule(r){7-9} \cmidrule(r){10-12} \cmidrule(r){13-15}
Method & Length & Exp. & $\mu$ & $\sigma$ & $n$\tnote{i} & $\mu$ & $\sigma$ & $n$ & $\mu$ & $\sigma$ & $n$ & $\mu$ & $\sigma$ & $n$\\
\hline
\multicolumn{15}{l}{\textbf{DIF in $\boldsymbol{b}$}}\\

\multicolumn{15}{l}{\hspace{0.5em}\textbf{for 20\% items}}\\

\hspace{1em}MLE & 25& .20 & .255 & .160 & 4 & .250 & .192 & 4 & .250 & .172 & 4 & .242 & .184 & 4\\
\hspace{1em} &  & .33 & .355 & .134 & 2 & .380 & .099 & 2 & .360 & .085 & 2 & .385 & .092 & 2\\
\hspace{1em} & 35 & .20 & .123 & .082 & 7 & .130 & .068 & 7 & .116 & .050 & 7 & .133 & .065 & 7\\
\hspace{1em} &  & .33 & .213 & .254 & 18 & .213 & .257 & 18 & .209 & .239 & 18 & .214 & .260 & 18\\

\hspace{1em}EAP & 25 & .20 & .285 & .170 & 4 & .292 & .160 & 4 & .272 & .156 & 4 & .283 & .152 & 4\\
\hspace{1em} &  & .33 & .315 & .177 & 2 & .305 & .035 & 2 & .305 & .021 & 2 & .285 & .092 & 2\\
\hspace{1em} & 35 & .20 & .144 & .060 & 7 & .147 & .094 & 7 & .133 & .083 & 7 & .154 & .092 & 7\\
\hspace{1em} &  & .33 & .270 & .276 & 15 & .276 & .285 & 15 & .274 & .271 & 15 & .275 & .284 & 15\\

\multicolumn{15}{l}{\hspace{0.5em}\textbf{for 40\% items}}\\

\hspace{1em}MLE & 25& .20 & .265 & .276 & 2 & .270 & .325 & 2 & .260 & .297 & 2 & .260 & .311 & 2\\
\hspace{1em} &  & .33 & .414 & . & 1 & .444 & . & 1 & .242 & . & 1 & .444 & . & 1\\
\hspace{1em} & 35 & .20 & .190 & .117 & 4 & .222 & .122 & 4 & .182 & .096 & 4 & .215 & .130 & 4\\
\hspace{1em} &  & .33 & .367 & .339 & 7 & .381 & .340 & 7 & .359 & .329 & 7 & .379 & .342 & 7\\

\hspace{1em}EAP & 25& .20 & .390 & .198 & 2 & .400 & .170 & 2 & .400 & .170 & 2 & .420 & .170 & 2\\
\hspace{1em} &  & .33 & .465 & . & 1 & .485 & . & 1 & .293 & . & 1 & .475 & . & 1\\
\hspace{1em} & 35 & .20 & .196 & .111 & 5 & .196 & .127 & 5 & .186 & .113 & 5 & .184 & .130 & 5\\
\hspace{1em} &  & .33 & .348 & .317 & 8 & .341 & .341 & 8 & .326 & .332 & 8 & .341 & .337 & 8\\
\hline
\multicolumn{15}{l}{\textbf{DIF in $\boldsymbol{a}$}}\\

\multicolumn{15}{l}{\hspace{0.5em}\textbf{for 20\% items}}\\

\hspace{1em}MLE & 25& .20 & .126 & .127 & 5 & .130 & .122 & 5 & .154 & .153 & 5 & .126 & .113 & 5\\
\hspace{1em} &  & .33 & .025 & .007 & 2 & .070 & .014 & 2 & .070 & .000 & 2 & .060 & .014 & 2\\
\hspace{1em} & 35 & .20 & .147 & .084 & 6 & .113 & .057 & 6 & .122 & .050 & 6 & .112 & .066 & 6\\
\hspace{1em} &  & .33 & .099 & .096 & 18 & .119 & .119 & 18 & .118 & .102 & 18 & .107 & .115 & 18\\

\hspace{1em}EAP & 25 & .20 & .180 & .141 & 6 & .183 & .155 & 6 & .205 & .163 & 6 & .190 & .167 & 6\\
\hspace{1em} &  & .33 & .060 & .057 & 2 & .095 & .021 & 2 & .090 & .028 & 2 & .065 & .021 & 2\\
\hspace{1em} & 35 & .20 & .137 & .065 & 7 & .127 & .081 & 7 & .117 & .072 & 7 & .133 & .079 & 7\\
\hspace{1em} &  & .33 & .099 & .099 & 16 & .124 & .128 & 16 & .141 & .124 & 16 & .111 & .121 & 16\\

\multicolumn{15}{l}{\hspace{0.5em}\textbf{for 40\% items}}\\

\hspace{1em}MLE & 25 & .20 & .090 & . & 1 & .110 & . & 1 & .120 & . & 1 & .100 & . & 1\\
\hspace{1em} & & .33 & .121 & . & 1 & .030 & . & 1 & .071 & . & 1 & .030 & . & 1\\
\hspace{1em} & 35 & .20 & .095 & .038 & 4 & .098 & .062 & 4 & .095 & .070 & 4 & .092 & .061 & 4\\
\hspace{1em} &  & .33 & .131 & .178 & 7 & .166 & .193 & 7 & .133 & .120 & 7 & .150 & .187 & 7\\

\hspace{1em}EAP & 25& .20 & .080 & . & 1 & .080 & . & 1 & .080 & . & 1 & .070 & . & 1\\
\hspace{1em} &  & .33 & .101 & . & 1 & .051 & . & 1 & .091 & . & 1 & .081 & . & 1\\
\hspace{1em} & 35 & .20 & .110 & .087 & 5 & .094 & .051 & 5 & .082 & .029 & 5 & .092 & .045 & 5\\
\hspace{1em} &  & .33 & .072 & .039 & 6 & .117 & .080 & 6 & .137 & .078 & 6 & .082 & .059 & 6\\

\hline
\end{tabular}
\begin{tablenotes}
\vspace{4pt}
        \textit{Note}. \textsuperscript{i} Effective item count across replications in which all candidate models converged; replications with non‑convergent models were excluded to ensure comparability.
    \end{tablenotes}
\end{threeparttable}
\end{table}

\begin{table}[htbp]
\footnotesize
\centering
\def\arraystretch{0.6} 
\begin{threeparttable}  
\caption{\label{tab:chp4_wDIF_power}Item-Level Statistical Power (Study 2, 100 Replications)}
\begin{tabular}[t]{lllrrrrrrrrrrrr}
\hline
\multicolumn{1}{c}{ } & \multicolumn{1}{c}{ } & \multicolumn{1}{c}{ } & \multicolumn{12}{c}{Models} \\
\multicolumn{1}{c}{Est.} & \multicolumn{1}{c}{Test} & \multicolumn{1}{c}{Max.} & \multicolumn{3}{c}{M6} & \multicolumn{3}{c}{S1} & \multicolumn{3}{c}{S2} & \multicolumn{3}{c}{S3} \\
\cmidrule(r){4-6} \cmidrule(r){7-9} \cmidrule(r){10-12} \cmidrule(r){13-15}
Method & Length & Exp. & $\mu$ & $\sigma$ & $n$\tnote{i} & $\mu$ & $\sigma$ & $n$ & $\mu$ & $\sigma$ & $n$ & $\mu$ & $\sigma$ & $n$\\
\hline
\multicolumn{15}{l}{\textbf{DIF in $\boldsymbol{b}$}}\\

\multicolumn{15}{l}{\hspace{0.5em}\textbf{for 20\% items}}\\

\hspace{1em}MLE & 25 & .20 & .397 & .223 & 4 & .398 & .203 & 4 & .392 & .171 & 4 & .390 & .223 & 4\\
\hspace{1em} &  & .33 & .290 & .014 & 2 & .330 & .000 & 2 & .305 & .035 & 2 & .305 & .007 & 2\\
\hspace{1em} & 35 & .20 & .227 & .244 & 7 & .251 & .240 & 7 & .207 & .211 & 7 & .261 & .247 & 7\\
\hspace{1em} &  & .33 & .163 & .158 & 12 & .150 & .148 & 12 & .152 & .138 & 12 & .148 & .142 & 12\\

\hspace{1em}EAP & 25 & .20 & .373 & .221 & 4 & .358 & .299 & 4 & .328 & .275 & 4 & .350 & .300 & 4\\
\hspace{1em} &  & .33 & .255 & .064 & 2 & .250 & .028 & 2 & .265 & .035 & 2 & .250 & .014 & 2\\
\hspace{1em} & 35 & .20 & .115 & .038 & 7 & .130 & .070 & 7 & .112 & .049 & 7 & .127 & .073 & 7\\
\hspace{1em} & & .33 & .214 & .281 & 13 & .228 & .269 & 13 & .228 & .261 & 13 & .228 & .278 & 13\\

\multicolumn{15}{l}{\hspace{0.5em}\textbf{for 40\% items}}\\

\hspace{1em}MLE & 25 & .20 & .247 & .194 & 2 & .203 & .194 & 2 & .170 & .132 & 2 & .214 & .179 & 2\\
\hspace{1em} &  & .33 & .404 & . & 1 & .427 & . & 1 & .404 & . & 1 & .438 & . & 1\\
\hspace{1em} & 35 & .20 & .065 & . & 1 & .097 & . & 1 & .075 & . & 1 & .086 & . & \vphantom{1} 1\\
\hspace{1em} &  & .33 & .745 & .276 & 2 & .735 & .276 & 2 & .720 & .240 & 2 & .740 & .283 & 2\\

\hspace{1em}EAP & 25& .20 & .424 & . & 1 & .413 & . & 1 & .348 & . & 1 & .413 & . & 1\\
\hspace{1em} &  & .33 & .360 & . & 1 & .337 & . & 1 & .303 & . & 1 & .337 & . & 1\\
\hspace{1em} & 35 & .20 & .032 & . & 1 & .086 & . & 1 & .097 & . & 1 & .065 & . & 1\\
\hspace{1em} &  & .33 & .740 & .283 & 2 & .755 & .276 & 2 & .740 & .269 & 2 & .755 & .262 & 2\\
\hline
\multicolumn{15}{l}{\textbf{DIF in $\boldsymbol{a}$}}\\

\multicolumn{15}{l}{\hspace{0.5em}\textbf{for 20\% items}}\\

\hspace{1em}MLE & 25& .20 & .338 & .307 & 5 & .344 & .266 & 5 & .362 & .280 & 5 & .334 & .280 & 5\\
\hspace{1em} &  & .33 & .140 & .127 & 2 & .165 & .092 & 2 & .130 & .057 & 2 & .155 & .106 & 2\\
\hspace{1em} & 35 & .20 & .103 & .087 & 6 & .092 & .083 & 6 & .075 & .054 & 6 & .092 & .079 & 6\\
\hspace{1em} &  & .33 & .057 & .037 & 12 & .082 & .064 & 12 & .085 & .058 & 12 & .072 & .060 & 12\\

\hspace{1em}EAP & 25& .20 & .359 & .264 & 6 & .347 & .295 & 6 & .304 & .271 & 6 & .339 & .301 & 6\\
\hspace{1em} &  & .33 & .100 & .099 & 2 & .125 & .021 & 2 & .135 & .035 & 2 & .105 & .049 & 2\\
\hspace{1em} & 35 & .20 & .138 & .082 & 7 & .129 & .104 & 7 & .122 & .073 & 7 & .143 & .101 & 7\\
\hspace{1em} &  & .33 & .081 & .058 & 15 & .090 & .066 & 15 & .092 & .081 & 15 & .084 & .061 & 15\\

\multicolumn{15}{l}{\hspace{0.5em}\textbf{for 40\% items}}\\

\hspace{1em}MLE & 25 & .20 & .076 & . & 1 & .261 & . & 1 & .217 & . & 1 & .250 & . & 1\\
\hspace{1em} &  & .33 & .034 & . & 1 & .090 & . & 1 & .135 & . & 1 & .090 & . & 1\\
\hspace{1em} & 35 & .20 & .108 & . & 1 & .129 & . & 1 & .065 & . & 1 & .140 & . & 1\\
\hspace{1em} &  & .33 & .070 & .014 & 2 & .120 & .071 & 2 & .105 & .064 & 2 & .135 & .064 & 2\\

\hspace{1em}EAP & 25& .20 & .109 & .077 & 2 & .158 & .085 & 2 & .179 & .115 & 2 & .136 & .085 & 2\\
\hspace{1em} &  & .33 & .011 & . & 1 & .011 & . & 1 & .034 & . & 1 & .011 & . & 1\\
\hspace{1em} & 35 & .20 & .065 & . & 1 & .097 & . & 1 & .075 & . & 1 & .086 & . & 1\\
\hspace{1em} &  & .33 & .110 & . & 1 & .070 & . & 1 & .090 & . & 1 & .050 & . & 1\\

\hline
\end{tabular}
\begin{tablenotes}
\vspace{4pt}
        \textit{Note}. \textsuperscript{i} Effective item count across replications in which all candidate models converged; replications with non‑convergent models were excluded to ensure comparability.
    \end{tablenotes}
\end{threeparttable}
\end{table}

\textbf{Interpreting the role of convergence (effective item counts $n$).} As mentioned before, results were computed only from replications in which all candidate models converged. We reported convergence here in Tables \ref{tab:chp4_nDIF_typ1}--\ref{tab:chp4_wDIF_power} using the number of administered items after excluding replications in which any candidate model failed to converge, or effective item count $n$. Across the simulation conditions, nonconvergence was strongly associated with DIF contamination in the data. Specifically, the effective item counts were dramatically smaller in Study 2---frequently in the single digits (e.g., $n=$ 1--7 in many 25-item conditions and $n\approx$ 12--18 in several 35-item, 20\%-DIF conditions)---indicating that joint convergence across all candidate models was achieved for only a small subset of administered items once DIF contamination was introduced. Also, longer tests (35 items) tended to yield larger $n$ than shorter tests (25 items) under several DIF scenarios (e.g., DIF in $b$ at 20\% with MLE/EAP and exposure .33), whereas shorter tests and/or higher DIF prevalence (40\%) were associated with the smallest $n$ values (often $n=$ 1--2), consistent with the idea that longer tests provide more stable ability estimation and model fitting. In addition, $n$ varied systematically as DIF prevalence changed (20\% vs. 40\%), indicating that nonconvergence was not random noise but condition-dependent, and that the severity of contamination and the resulting instability in model estimation made a difference on $n$. Consequently, low power in some Study 2 cells should be interpreted in conjunction with $n$: In settings where the effective item count is small, both power and Type-I error estimates are necessarily less stable and also reflect the practical difficulty of fitting the models under those CAT DIF configurations. This nonconvergence-power association implied that inferential performance of models is intertwined with convergence reliability in complex CAT DIF analyses.
\\
Taken together, we reached the following conclusions from Monte Carlo simulation results. First, under DIF-free conditions, the proposed multilevel model M6 offered the most reliable protection against false-positive DIF findings in CAT data by accounting for provisional-ability-induced dependency. Second, when DIF was present, power was strongly driven by CAT design features such as test length and item exposure. M6 tracked closely with the single-level alternatives and, in several low-information settings (i.e., with shorter tests or lower item exposure), showed slightly higher power. This pattern suggested that accounting for provisional-ability-induced dependency can provide a modest advantage when information is limited. Lastly, convergence behavior varied systematically across conditions and directly affected the effective information used for inference. This reality underscored the need to consider convergence reliability as part of practical method performance in CAT DIF applications.

\section{Discussion}

In this study, we demonstrated that structural dependencies exist item‑adaptive CAT data, as provisional ability estimates drive item selection. Because these estimates depend on previously administered items, CAT responses violate the independence assumptions underlying conventional single‑level DIF methods. By conceptualizing examinees as nested within provisional‑ability intervals, we showed that CAT‑induced dependency can be modeled directly using multilevel logistic regression without modifying the operational CAT algorithm. This framing provided a principled statistical foundation for DIF analysis in adaptive testing environments. We proposed a two-level model for analyzing DIF in CAT data because it accounts for the nested structure of item responses, which traditional single-level methods like LR and MH cannot achieve. 
\\
We evaluated our proposed two‑level model using both a numeric example and Monte Carlo simulations. In the numeric example, we illustrated how DIF patterns shown in item characteristic curves varied across provisional‑ability intervals and demonstrated that accounting for this structure can materially affect DIF inference through model comparisons. In Monte Carlo simulations, we showed that the two‑level model that stood out in the numeric example provided the strongest protection against false‑positive DIF under DIF‑free conditions and that it performed comparably to single‑level methods when DIF was present, with a modest advantage in low‑information settings such as shorter tests and lower exposure rates. These findings indicated that explicitly accounting for CAT‑induced dependency via multilevel modeling holds promise for improving the robustness of DIF inference with CAT data.
\\
In this study, we deliberately based all the simulations on item response data generated under a CAT algorithm consistent with operational practice, rather than simplifying the data‑generating process. This design ensured that models were evaluated under realistic combinations of adaptive item selection criteria. However, it also introduced additional variability and estimation challenges that can attenuate power, particularly under high DIF prevalence conditions. As a result, reduced power in some conditions should be interpreted as a consequence of model evaluation under structurally complex CAT settings.
\\
An important insight from Monte Carlo simulations is that convergence behavior varied systematically across CAT and DIF conditions and directly affected the effective information available for inference. Under DIF‑free conditions, all models generally converged across a wide range of CAT designs, yielding stable item‑level summaries. In contrast, under DIF‑contaminated conditions---particularly with shorter tests, higher DIF prevalence, or Parameter-$a$‑based DIF---the effective item count was substantially reduced, which could partly explain the low statistical power observed in Study 2. This pattern reflected the increased estimation difficulty introduced by DIF contamination in CAT settings, and underscored that convergence reliability is an inherent component of practical DIF modeling in CAT rather than a purely numerical artifact.
\\
These findings have two implications. First, the reported Type‑I error and power estimates should be understood as conditional on joint model convergence, reflecting the information that would be available to practitioners in applied settings. Second, the systematic variation in item counts across conditions underscored that convergence behavior itself is a meaningful component of model performance in CAT DIF analysis. Our results highlighted the need for future research to jointly evaluate convergence probability and inferential accuracy. Designing simulation studies that separate the two would be a promising direction for extending the present work.
\\
Several directions for future research follow from this work. Expanding item pools might increase the effective item count under DIF‑contaminated conditions, particularly when DIF prevalence is high. Incorporating item parameters directly into the multilevel structure—for example, via crossed random‑effects models for ability intervals and items—may further improve DIF estimation, although such extensions raise additional convergence challenges. Bayesian estimation offers a promising avenue for addressing these challenges by stabilizing parameter estimation in complex hierarchical models. Further investigations could also explore how CAT‑specific factors such as examinee sample size (e.g., 10,000 vs. 5,000), item selection algorithms (e.g., STA vs. WPM), exposure control strategies (e.g., Fade Away Strategy, Sympson Hetter Strategy), and DIF magnitude (e.g., .4 vs .1) interact with multilevel modeling choices (e.g., ICC, the number of provisional ability intervals, and the sample size for each interval) 
to influence model performance. More broadly, future studies should explicitly examine convergence probability alongside inferential accuracy, as the present results---conditional on joint model convergence---cannot fully disentangle Type‑I error and power from convergence reliability. Examining these two aspects jointly would provide a more complete evaluation of DIF model performance in CAT and help clarify the trade‑offs between statistical rigor and operational feasibility.
\\
For practitioners conducting DIF analyses with CAT data, we want to highlight the importance of accounting for CAT‑induced dependency when evaluating DIF. Standard single‑level DIF methods that condition on total scores or final ability estimates may overflag items or miss true DIF because they ignore the structural dependencies in the CAT data. Multilevel DIF models that incorporate provisional‑ability information offer a promising alternative, particularly for protecting against false‑positive DIF. At the same time, practitioners should recognize that DIF detection remains sensitive to CAT design features such as test length and exposure control, and that convergence behavior is an integral part of practical method performance when working with complex CAT data.

\section{Conclusion}
Through this study, we advanced DIF methodology for CAT by proposing a two-level logistic model to improve DIF inference with CAT data. By modeling examinees as nested within provisional‑ability intervals, we demonstrated that our proposed multilevel framework can provide a principled way to account for CAT‑induced dependency without modifying the operational CAT algorithm. Using a numeric example and Monte Carlo simulation studies, we unveiled that explicitly modeling this nested structure improves protection against spurious DIF under DIF‑free conditions and yields competitive performance when DIF is present, particularly in low‑information conditions. Meanwhile, we highlighted that DIF detection in CAT is inherently constrained by design features and convergence behavior and that convergence reliability shall be considered as part of practical model performance. Overall, we illustrated the promise of multilevel modeling for improving the robustness of DIF inference in CAT and provided a foundation for future research that jointly evaluates inferential accuracy, convergence, and operational feasibility.

\printbibliography

@thesis{chen_modeling_2023,
  title = {Modeling Item Bias in Fixed-Item Tests and Computerized Adaptive Tests},
  author = {Chen, Dandan},
  date = {2023},
  institution = {{University of Illinois at Urbana-Champaign}},
  location = {{Urbana, IL}},
  url = {https://www.ideals.illinois.edu/items/127410}
}

@book{fishbein_timss_2021,
  title = {{{TIMSS}} 2019 User Guide for the International Database},
  author = {Fishbein, B and Foy, P and Yin, L},
  date = {2021},
  edition = {2},
  publisher = {{TIMSS \& PIRLS International Study Center}},
  location = {{Chestnut Hill, MA}}
}

@book{hambleton_fundamentals_1991,
  title = {Fundamentals of Item Response Theory},
  author = {Hambleton, Ronald K. and Swaminathan, Hariharan and Rogers, H. Jane},
  date = {1991},
  eprint = {cmJU9SHCzecC},
  eprinttype = {googlebooks},
  publisher = {{SAGE Publications, Inc}},
  location = {{Newbury Park, CA}},
  isbn = {978-0-8039-3647-8},
  langid = {english},
  pagetotal = {192}
}

@report{holland_alternate_1985,
  title = {An Alternate Definition of the {{ETS}} Delta Scale of Item Difficulty},
  author = {Holland, P and Thayer, Dorothy T.},
  date = {1985-10},
  series = {{{ETS Research Report Series}}},
  number = {85-43},
  pages = {i--10},
  institution = {{Educational Testing Service}},
  location = {{Princeton, NJ}},
  url = {https://onlinelibrary.wiley.com/doi/10.1002/j.2330-8516.1985.tb00128.x},
  urldate = {2022-01-22},
  langid = {english}
}

@book{mullis_timss_2017,
  title = {{{TIMSS}} 2019 Assessment Frameworks},
  editor = {Mullis, I and Martin, M},
  date = {2017},
  publisher = {{TIMSS \& PIRLS International Study Center}},
  location = {{Boston, MA}},
  url = {http://timssandpirls.bc.edu/timss2019/frameworks/}
}

@article{penfield_assessing_2001,
  title = {Assessing Differential Item Functioning among Multiple Groups: {{A}} Comparison of Three {{Mantel-Haenszel}} Procedures},
  shorttitle = {Assessing {{Differential Item Functioning Among Multiple Groups}}},
  author = {Penfield, Randall D.},
  date = {2001-07-01},
  journaltitle = {Applied Measurement in Education},
  volume = {14},
  number = {3},
  pages = {235--259},
  publisher = {{Routledge}},
  issn = {0895-7347},
  doi = {10.1207/S15324818AME1403_3},
  url = {https://doi.org/10.1207/S15324818AME1403_3},
  urldate = {2022-01-22}
}

@software{r_core_team_r_2021,
  title = {R: {{A}} Language and Environment for Statistical Computing},
  author = {{R Core Team}},
  date = {2021},
  location = {{Vienna, Austria}},
  url = {https://www.R-project.org/},
  organization = {{R Foundation for Statistical Computing}},
  version = {Version 4.0.4}
}

@article{shealy_model-based_1993,
  title = {A Model-Based Standardization Approach That Separates True Bias/{{DIF}} from Group Ability Differences and Detects Test Bias/{{DTF}} as Well as Item Bias/{{DIF}}},
  author = {Shealy, R and Stout, W},
  date = {1993-06},
  journaltitle = {Psychometrika},
  shortjournal = {Psychometrika},
  volume = {58},
  number = {2},
  pages = {159--194},
  issn = {0033-3123, 1860-0980},
  doi = {10.1007/BF02294572},
  url = {http://link.springer.com/10.1007/BF02294572},
  urldate = {2022-01-21},
  langid = {english}
}

@article{svetina_detecting_2014,
  title = {Detecting Differential Item Functioning Using Generalized Logistic Regression in the Context of Large-Scale Assessments},
  author = {Svetina, Dubravka and Rutkowski, Leslie},
  date = {2014-12},
  journaltitle = {Large-scale Assessments in Education},
  shortjournal = {Large-scale Assess Educ},
  volume = {2},
  number = {1},
  pages = {4},
  issn = {2196-0739},
  doi = {10.1186/s40536-014-0004-5},
  url = {https://largescaleassessmentsineducation.springeropen.com/articles/10.1186/s40536-014-0004-5},
  urldate = {2022-01-21},
  langid = {english}
}

@article{swaminathan_detecting_1990,
  title = {Detecting Differential Item Functioning Using Logistic Regression Procedures},
  author = {Swaminathan, Hariharan and Rogers, H. Jane},
  date = {1990-12},
  journaltitle = {Journal of Educational Measurement},
  shortjournal = {J Educational Measurement},
  volume = {27},
  number = {4},
  pages = {361--370},
  issn = {0022-0655, 1745-3984},
  doi = {10.1111/j.1745-3984.1990.tb00754.x},
  url = {https://onlinelibrary.wiley.com/doi/10.1111/j.1745-3984.1990.tb00754.x},
  urldate = {2022-01-21},
  langid = {english}
}

@article{thissen_beyond_1986,
  title = {Beyond Group-Mean Differences: {{The}} Concept of Item Bias},
  shorttitle = {Beyond Group-Mean Differences},
  author = {Thissen, D and Steinberg, Lynne and Gerrard, Meg},
  date = {1986-01},
  journaltitle = {Psychological Bulletin},
  volume = {99},
  number = {1},
  pages = {118--128},
  publisher = {{American Psychological Association}},
  issn = {0033-2909},
  doi = {http://dx.doi.org.proxy2.library.illinois.edu/10.1037/0033-2909.99.1.118},
  url = {http://www.proquest.com/docview/614321833/abstract/91A6A05F584B4069PQ/1},
  urldate = {2022-02-23},
  langid = {english},
  pagetotal = {118-128}
}

@article{woods_langer-improved_2013,
  title = {The {{Langer-improved Wald}} Test for {{DIF}} Testing with Multiple Groups: {{Evaluation}} and Comparison to Two-Group {{IRT}}},
  shorttitle = {The {{Langer-Improved Wald Test}} for {{DIF Testing With Multiple Groups}}},
  author = {Woods, Carol M. and Cai, Li and Wang, Mian},
  date = {2013-06},
  journaltitle = {Educational and Psychological Measurement},
  shortjournal = {Educational and Psychological Measurement},
  volume = {73},
  number = {3},
  pages = {532--547},
  issn = {0013-1644, 1552-3888},
  doi = {10.1177/0013164412464875},
  url = {http://journals.sagepub.com/doi/10.1177/0013164412464875},
  urldate = {2022-01-21},
  langid = {english}
}

@article{ali_sample_2019,
  title = {Sample Size Issues in Multilevel Logistic Regression Models},
  author = {Ali, Amjad and Ali, Sabz and Khan, Sajjad Ahmad and Khan, Dost Muhammad and Abbas, Kamran and Khalil, Alamgir and Manzoor, Sadaf and Khalil, Umair},
  date = {2019-11-22},
  journaltitle = {PLOS ONE},
  shortjournal = {PLOS ONE},
  volume = {14},
  number = {11},
  pages = {e0225427},
  publisher = {{Public Library of Science}},
  issn = {1932-6203},
  doi = {10.1371/journal.pone.0225427},
  url = {https://journals.plos.org/plosone/article?id=10.1371/journal.pone.0225427},
  urldate = {2023-02-06},
  langid = {english}
}

@software{bates_linear_2022-1,
  title = {Linear Mixed-Effects Models Using '{{Eigen}}' and {{S4}}},
  author = {Bates, D and Maechler, M and Bolker, B and Walker, S},
  date = {2022},
  url = {https://cran.r-project.org/web/packages/lme4/lme4.pdf},
  editora = {Christensen, R and Singmann, H and Dai, B and Scheipl, F and Grothendieck, G and Green, P and Fox, J and Bauer, A and Krivitsky, P},
  editoratype = {collaborator},
  version = {1.1-31}
}

@article{bock_item_1988,
  title = {Item Pool Maintenance in the Presence of Item Parameter Drift},
  author = {Bock, R and Murakl, Eiji and Pfeiffenberger, Will},
  date = {1988},
  journaltitle = {Journal of Educational Measurement},
  volume = {25},
  number = {4},
  pages = {275--285},
  issn = {1745-3984},
  doi = {10.1111/j.1745-3984.1988.tb00308.x},
  url = {https://onlinelibrary.wiley.com/doi/abs/10.1111/j.1745-3984.1988.tb00308.x},
  urldate = {2023-01-19},
  langid = {english}
}

@article{chang_-stratified_1999,
  title = {A-Stratified Multistage Computerized Adaptive Testing},
  author = {Chang, H and Ying, Zhiliang},
  date = {1999-09},
  journaltitle = {Applied Psychological Measurement},
  shortjournal = {Applied Psychological Measurement},
  volume = {23},
  number = {3},
  pages = {211--222},
  issn = {0146-6216, 1552-3497},
  doi = {10.1177/01466219922031338},
  url = {http://journals.sagepub.com/doi/10.1177/01466219922031338},
  urldate = {2022-09-27},
  langid = {english}
}

@article{chang_comparative_2003,
  title = {A Comparative Study of Item Exposure Control Methods in Computerized Adaptive Testing},
  author = {Chang, S and Ansley, Timothy N.},
  date = {2003},
  journaltitle = {Journal of Educational Measurement},
  volume = {40},
  number = {1},
  eprint = {1435055},
  eprinttype = {jstor},
  pages = {71--103},
  publisher = {{[National Council on Measurement in Education, Wiley]}},
  issn = {0022-0655},
  url = {http://www.jstor.org/stable/1435055},
  urldate = {2022-12-15}
}

@article{chang_nonlinear_2009,
  title = {Nonlinear Sequential Designs for Logistic Item Response Theory Models with Applications to Computerized Adaptive Tests},
  author = {Chang, H and Ying, Zhiliang},
  date = {2009},
  journaltitle = {The Annals of Statistics},
  volume = {37},
  number = {3},
  eprint = {30243674},
  eprinttype = {jstor},
  pages = {1466--1488},
  publisher = {{Institute of Mathematical Statistics}},
  issn = {0090-5364},
  url = {https://www.jstor.org/stable/30243674},
  urldate = {2023-02-20}
}

@article{chang_psychometrics_2015,
  title = {Psychometrics behind Computerized Adaptive Testing},
  author = {Chang, H},
  date = {2015-03-01},
  journaltitle = {Psychometrika},
  shortjournal = {Psychometrika},
  volume = {80},
  number = {1},
  pages = {1--20},
  issn = {1860-0980},
  doi = {10.1007/s11336-014-9401-5},
  url = {https://doi.org/10.1007/s11336-014-9401-5},
  urldate = {2023-02-20},
  langid = {english}
}

@unpublished{chu_detecting_2012,
title= {Detecting Directional DIF Using CATSIB with Impact Present},
author = {Chu, Man-Wai and Lai, Hollis and Wang, Xian},
year = {2012},
note= {Conference presentation},
addendum     = {Annual Meeting of National Council on Measurement in Education, Vancouver, British Columbia, Canada.}
}

@article{dorman_effect_2008,
  title = {The Effect of Clustering on Statistical Tests: An Illustration Using Classroom Environment Data},
  shorttitle = {The Effect of Clustering on Statistical Tests},
  author = {Dorman, Jeffrey Paul},
  date = {2008-08-01},
  journaltitle = {Educational Psychology},
  volume = {28},
  number = {5},
  pages = {583--595},
  publisher = {{Routledge}},
  issn = {0144-3410},
  doi = {10.1080/01443410801954201},
  url = {https://doi.org/10.1080/01443410801954201},
  urldate = {2022-06-23}
}

@book{gelman_data_2007,
  title = {Data Analysis Using Regression and Multilevel/Hierarchical Models},
  author = {Gelman, Andrew and Hill, Jennifer},
  date = {2007},
  publisher = {{Cambridge University Press}},
  location = {{New York, NY}}
}

@article{goldstein_partitioning_2002,
  title = {Partitioning Variation in Multilevel Models},
  author = {Goldstein, Harvey and Browne, William and Rasbash, Jon},
  date = {2002-12-02},
  journaltitle = {Understanding Statistics},
  volume = {1},
  number = {4},
  pages = {223--231},
  publisher = {{Routledge}},
  issn = {1534-844X},
  doi = {10.1207/S15328031US0104_02},
  url = {https://doi.org/10.1207/S15328031US0104_02},
  urldate = {2022-02-28}
}

@article{hedges_intraclass_2007,
  title = {Intraclass Correlation Values for Planning Group-Randomized Trials in Education},
  author = {Hedges, Larry V. and Hedberg, E. C.},
  date = {2007-03-01},
  journaltitle = {Educational Evaluation and Policy Analysis},
  shortjournal = {Educational Evaluation and Policy Analysis},
  volume = {29},
  number = {1},
  pages = {60--87},
  publisher = {{American Educational Research Association}},
  issn = {0162-3737},
  doi = {10.3102/0162373707299706},
  url = {https://doi.org/10.3102/0162373707299706},
  urldate = {2022-02-28}
}

@article{kingsbury_procedures_1989-2,
  title = {Procedures for Selecting Items for Computerized Adaptive Tests},
  author = {Kingsbury, G. Gage and Zara, Anthony R.},
  date = {1989-10},
  journaltitle = {Applied Measurement in Education},
  volume = {2},
  number = {4},
  pages = {359},
  publisher = {{Taylor \& Francis Ltd}},
  issn = {08957347},
  doi = {10.1207/s15324818ame0204_6},
  url = {https://www.tandfonline.com/doi/abs/10.1207/s15324818ame0204_6},
  urldate = {2022-10-14}
}

@article{lei_comparing_2006,
  title = {Comparing Methods of Assessing Differential Item Functioning in a Computerized Adaptive Testing Environment},
  author = {Lei, Pui-Wa and Chen, Shu-Ying and Yu, Lan},
  date = {2006-09},
  journaltitle = {Journal of Educational Measurement},
  shortjournal = {J Educational Measurement},
  volume = {43},
  number = {3},
  pages = {245--264},
  issn = {0022-0655, 1745-3984},
  doi = {10.1111/j.1745-3984.2006.00015.x},
  url = {https://onlinelibrary.wiley.com/doi/10.1111/j.1745-3984.2006.00015.x},
  urldate = {2022-08-09},
  langid = {english}
}

@article{mislevy_does_2000,
  title = {Does Adaptive Testing Violate Local Independence?},
  author = {Mislevy, Robert and Chang, Hua-Hua},
  date = {2000-06-01},
  journaltitle = {Psychometrika},
  shortjournal = {Psychometrika},
  volume = {65},
  number = {2},
  pages = {149--156},
  issn = {1860-0980},
  doi = {10.1007/BF02294370},
  url = {https://doi.org/10.1007/BF02294370},
  urldate = {2022-12-12},
  langid = {english}
}

@article{moineddin_simulation_2007,
  title = {A Simulation Study of Sample Size for Multilevel Logistic Regression Models},
  author = {Moineddin, Rahim and Matheson, Flora I and Glazier, Richard H},
  date = {2007-12},
  journaltitle = {BMC Medical Research Methodology},
  shortjournal = {BMC Med Res Methodol},
  volume = {7},
  number = {34},
  pages = {1--10},
  issn = {1471-2288},
  doi = {10.1186/1471-2288-7-34},
  url = {https://bmcmedresmethodol.biomedcentral.com/articles/10.1186/1471-2288-7-34},
  urldate = {2022-01-22},
  langid = {english}
}

@report{nandakumar_catsib_2001,
  title = {{{CATSIB}}: {{A}} Modified {{SIBTEST}} Procedure to Detect Differential Item Functioning in Computerized Adaptive Tests},
  author = {Nandakumar, R and Roussos, L},
  date = {2001},
  series = {{{LSAC Research Report Series}}},
  institution = {{Law School Admission Council, Inc}},
  url = {https://eric.ed.gov/?id=ED467821}
}

@book{raudenbush_hierarchical_2002,
  title = {Hierarchical Linear Models: {{Applications}} and Data Analysis Methods},
  author = {Raudenbush, S and Bryk, A},
  date = {2002},
  edition = {2},
  publisher = {{SAGE Publications, Inc}},
  location = {{Thousand Oaks, CA}}
}

@article{raudenbush_maximum_2000,
  title = {Maximum Likelihood for Generalized Linear Models with Nested Random Effects via High-Order, Multivariate {{Laplace}} Approximation},
  author = {Raudenbush, S and Yang, Meng-Li and Yosef, Matheos},
  date = {2000-03-01},
  journaltitle = {Journal of Computational and Graphical Statistics},
  volume = {9},
  number = {1},
  pages = {141--157},
  publisher = {{Taylor \& Francis}},
  issn = {1061-8600},
  doi = {10.1080/10618600.2000.10474870},
  url = {https://www.tandfonline.com/doi/abs/10.1080/10618600.2000.10474870},
  urldate = {2023-02-09}
}

@article{ren_continuous_2017-2,
  title = {Continuous Online Item Calibration: {{Parameter}} Recovery and Item Utilization},
  shorttitle = {Continuous {{Online Item Calibration}}},
  author = {Ren, Hao and van der Linden, Wim J. and Diao, Qi},
  date = {2017-06-01},
  journaltitle = {Psychometrika},
  shortjournal = {Psychometrika},
  volume = {82},
  number = {2},
  pages = {498--522},
  issn = {1860-0980},
  doi = {10.1007/s11336-017-9553-1},
  url = {https://doi.org/10.1007/s11336-017-9553-1},
  urldate = {2022-10-23},
  langid = {english}
}

@software{ren_pearson_2017,
  title = {Pearson {{CAT}} Simulation Program},
  author = {Ren, H and Martinez, J and Shin, D},
  date = {2017},
  publisher = {{Pearson, Inc}},
  version = {3.14.5}
}

@book{shadish_experimental_2002,
  title = {Experimental and Quasi-Experimental Designs for Generalized Causal Inference},
  author = {Shadish, William R. and Cook, Thomas D. and Campbell, Donald T.},
  date = {2002},
  edition = {2},
  publisher = {{Houghton Mifflin Company}},
  location = {{Boston}},
  isbn = {978-0-395-61556-0},
  langid = {english}
}

@unpublished{shin_comparison_2012,
title= {A comparison of three content balancing methods for fixed and variable length computerized adaptive tests},
author = {Shin, Chingwei David and Chien, Yuehmei and Way, Walter Denny},
year = {2012},
note= {Conference presentation},
addendum     = {Annual Meeting of National Council on Measurement in Education, Vancouver, British Columbia, Canada.}
}

@article{shin_conditional_2017,
  title = {Conditional randomesque method for item exposure control in CAT},
  author = {Shin, Chingwei David},
  date = {2017-09-01},
  journaltitle = {International Journal of Intelligent Technologies and Applied Statistics},
  volume = {10},
  number = {3},
   pages = {145--155},
  doi = {10.6148/IJITAS.2017.1003.02},
  langid = {english}
}

@report{shin_weighted_2009,
  title = {Weighted Penalty Model for Content Balancing in {{CATs}}},
  author = {Shin, Chingwei David and Chien, Y and Way, W and Swanson, L},
  date = {2009},
  institution = {{Pearson}},
  url = {http://images.pearsonclinical.com/images/tmrs/tmrs_rg/WeightedPenaltyModel.pdf},
  langid = {english}
}

@book{snijders_multilevel_2011,
  title = {Multilevel Analysis: {{An}} Introduction to Basic and Advanced Multilevel Modeling},
  shorttitle = {Multilevel {{Analysis}}},
  author = {Snijders, Tom A. B. and Bosker, R},
  date = {2011-11-04},
  edition = {2},
  publisher = {{SAGE Publications, Inc}},
  location = {{Thousand Oaks, CA}},
  isbn = {978-1-84920-201-5},
  langid = {english},
  pagetotal = {368}
}

@report{stocking_scale_1988,
  title = {Scale Drift in On-Line Calibration},
  author = {Stocking, Martha L.},
  date = {1988},
  series = {{{ETS Research Report Series}}},
  pages = {i-122},
  institution = {{Educational Testing Service}},
  location = {{Princeton, NJ}},
  url = {http://onlinelibrary.wiley.com/doi/abs/10.1002/j.2330-8516.1988.tb00284.x},
  urldate = {2022-10-31},
  langid = {english}
}

@article{stocking_three_1994,
  title = {Three Practical Issues for Modern Adaptive Testing Item Pools},
  author = {Stocking, Martha L.},
  date = {1994},
  journaltitle = {ETS Research Report Series},
  volume = {1994},
  number = {1},
  pages = {i-34},
  issn = {2330-8516},
  doi = {10.1002/j.2333-8504.1994.tb01578.x},
  url = {https://onlinelibrary.wiley.com/doi/abs/10.1002/j.2333-8504.1994.tb01578.x},
  urldate = {2023-01-21},
  langid = {english}
}

@article{van_der_linden_assembling_2006,
  title = {Assembling a Computerized Adaptive Testing Item Pool as a Set of Linear Tests},
  author = {family=Linden, given=W, prefix=van der, useprefix=true and Ariel, Adelaide and Veldkamp, Bernard P.},
  date = {2006},
  journaltitle = {Journal of Educational and Behavioral Statistics},
  volume = {31},
  number = {1},
  eprint = {3701289},
  eprinttype = {jstor},
  pages = {81--99},
  publisher = {{[American Educational Research Association, Sage Publications, Inc., American Statistical Association]}},
  issn = {1076-9986},
  url = {https://www.jstor.org/stable/3701289},
  urldate = {2023-01-21}
}

@incollection{van_der_linden_item_2000,
  title = {Item Selection and Ability Estimation in Adaptive Testing},
  booktitle = {Computerized Adaptive Testing: {{Theory}} and Practice},
  author = {family=Linden, given=W, prefix=van der, useprefix=true and Pashley, P},
  editor = {family=Linden, given=W, prefix=van der, useprefix=true and Glas, C},
  date = {2000},
  pages = {1--26},
  publisher = {{Kluwer Acadeic Publishers}},
  location = {{Norwell, MA}}
}

@article{van_der_linden_shadow-test_2020,
  title = {A Shadow-Test Approach to Adaptive Item Calibration},
  author = {family=Linden, given=W, prefix=van der, useprefix=true and Jiang, Bingnan},
  date = {2020-06-01},
  journaltitle = {Psychometrika},
  shortjournal = {Psychometrika},
  volume = {85},
  number = {2},
  pages = {301--321},
  issn = {1860-0980},
  doi = {10.1007/s11336-020-09703-8},
  url = {https://doi.org/10.1007/s11336-020-09703-8},
  urldate = {2022-10-19},
  langid = {english}
}

@article{vander_linden_optimal_2015,
  title = {Optimal {{Bayesian}} Adaptive Design for Test-Item Calibration},
  author = {family=Linden, given=W, prefix=van~der, useprefix=true and Ren, Hao},
  date = {2015-06-01},
  journaltitle = {Psychometrika},
  shortjournal = {Psychometrika},
  volume = {80},
  number = {2},
  pages = {263--288},
  issn = {1860-0980},
  doi = {10.1007/s11336-013-9391-8},
  url = {https://doi.org/10.1007/s11336-013-9391-8},
  urldate = {2022-10-23},
  langid = {english}
}

@article{veerkamp_detection_2000,
  title = {Detection of Known Items in Adaptive Testing with a Statistical Quality Control Method},
  author = {Veerkamp, Wim J. J. and Glas, Cees A. W.},
  date = {2000-12-01},
  journaltitle = {Journal of Educational and Behavioral Statistics},
  volume = {25},
  number = {4},
  pages = {373--389},
  publisher = {{American Educational Research Association}},
  issn = {1076-9986},
  doi = {10.3102/10769986025004373},
  url = {https://doi.org/10.3102/10769986025004373},
  urldate = {2023-01-19},
  langid = {english}
}

@book{wainer_computerized_2000,
  title = {Computerized Adaptive Testing: {{A}} Primer},
  author = {Wainer, H},
  date = {2000},
  edition = {2nd ed.},
  publisher = {{Lawrence Erlbaum Associates, Inc}}
}

@article{zwick_simulation_1994,
  title = {A Simulation Study of Methods for Assessing Differential Item Functioning in Computerized Adaptive Tests},
  author = {Zwick, Rebecca and Thayer, Dorothy T. and Wingersky, Marilyn},
  date = {1994-06-01},
  journaltitle = {Applied Psychological Measurement},
  shortjournal = {Applied Psychological Measurement},
  volume = {18},
  number = {2},
  pages = {121--140},
  publisher = {{SAGE Publications Inc}},
  issn = {0146-6216},
  doi = {10.1177/014662169401800203},
  url = {https://doi.org/10.1177/014662169401800203},
  urldate = {2022-08-09}
}

@article{burham_david_2004,
author = {Kenneth P. Burnham and David R. Anderson},
title ={Multimodel Inference: Understanding AIC and BIC in Model Selection},

journal = {Sociological Methods \& Research},
volume = {33},
number = {2},
pages = {261-304},
year = {2004},
doi = {10.1177/0049124104268644}
}

@article{lim_2024,
author = {Lim, Hwanggyu and Zhu, Danqi and Choe, Edison M. and Han, KyungT.},
title = {Detecting Differential Item Functioning among Multiple Groups Using {IRT} Residual {DIF} Framework},
journal = {Journal of Educational Measurement},
volume = {61},
number = {4},
pages = {656-681},
doi = {https://doi.org/10.1111/jedm.12415},
url = {https://onlinelibrary.wiley.com/doi/abs/10.1111/jedm.12415},
eprint = {https://onlinelibrary.wiley.com/doi/pdf/10.1111/jedm.12415},
year = {2024}
}

\end{document}